\documentclass[12pt,a4paper]{article}
\pdfoutput=1
\usepackage[utf8]{inputenc}
\usepackage{amsmath}
\usepackage{amsfonts} 
\usepackage{mathtools} 
\usepackage{amssymb} 
\usepackage{subfigure} 
\usepackage{graphicx}
\usepackage{xcolor}
\usepackage{cite} 
\usepackage{mathrsfs}
\usepackage{hyperref}

\newcommand{\abs}[1]{|#1|} 
\newcommand{\re}[1]{\text{Re}\left(#1\right)}
\newcommand{\im}[1]{\text{Im}\left(#1\right)}
\newcommand{\refEQ}[1]{eq.\,\eqref{#1}} 
\newcommand{\REFEQ}[1]{Eq.\,\eqref{#1}} 
\newcommand{\refEQS}[1]{eqs.\,\eqref{#1}} 
\newcommand{\Hc}{\text{H.c.}}

\newcommand{\SMHD}{\Phi}\newcommand{\SMHDd}{\Phi^\dagger}
\newcommand{\HD}[1]{\Phi_{#1}^{\phantom{\dagger}}}
\newcommand{\HDd}[1]{\Phi_{#1}^\dagger}

\newcommand{\HDc}[1]{\Phi_{#1}^{\phantom{\dagger}\!\!\!\ast}}
\newcommand{\HHD}[1]{H_{#1}^{\phantom{\dagger}}}
\newcommand{\HHDd}[1]{H_{#1}^\dagger}

\newcommand{\nHH}{{H}^0}
\newcommand{\nHR}{{R}^0}
\newcommand{\nHI}{{I}^0}
\newcommand{\nh}{{\rm h}}
\newcommand{\nhSM}{{\rm h_{SM}}}
\newcommand{\nH}{{\rm H}}
\newcommand{\nA}{{\rm A}}
\newcommand{\cH}{{\rm H}^\pm}

\newcommand{\cHp}{{\rm H}^+}
\newcommand{\mnh}{m_{\nh}}
\newcommand{\mnhSM}{m_{\nhSM}}
\newcommand{\mnH}{m_{\nH}}
\newcommand{\mnA}{m_{\nA}}
\newcommand{\mcH}{m_{\cH}}
\newcommand{\mNSc}{\mathcal M_0^2}

\newcommand{\cb}{c_\beta}
\renewcommand{\sb}{s_\beta}
\newcommand{\sbN}[1]{s_{#1 \beta}}
\newcommand{\cbN}[1]{c_{#1 \beta}}
\newcommand{\cbb}{c_{2\beta}}
\newcommand{\sbb}{s_{2\beta}}
\newcommand{\cba}{c_{\alpha\beta}}
\newcommand{\sba}{s_{\alpha\beta}}
\newcommand{\tb}{t_\beta}

\newcommand{\tbinv}{\tb^{-1}}
\newcommand{\tti}{\tb+\tbinv}

\newcommand{\VEV}[1]{\langle #1 \rangle}
\newcommand{\vev}[1]{v_{#1}}
\newcommand{\tCP}{\theta}
\newcommand{\stCP}{s_{\tCP}}
\newcommand{\sttCP}{s_{2\tCP}}
\newcommand{\ctCP}{c_{\tCP}}
\newcommand{\cttCP}{c_{2\tCP}}

\newcommand{\tinvtCP}{t_{\tCP}^{-1}}
\newcommand{\ROTmat}{\mathcal R}\newcommand{\ROTmatT}{\ROTmat^T}\newcommand{\ROTmatinv}{\ROTmat^{-1}}
\newcommand{\ROT}[1]{\ROTmat_{#1}}

\newcommand{\HbROT}{\mathcal R_{\beta}^{\phantom{T}}}\newcommand{\HbROTt}{\mathcal R_{\beta}^T}\newcommand{\HbROTinv}{\mathcal R_{\beta}^{-1}}

\newcommand{\Zn}[1]{\mathbb{Z}_{#1}}
\newcommand{\ZZ}{\Zn{2}}

\graphicspath{{Figs/}}

\begin{document}

\hfill\begin{minipage}[r]{0.3\textwidth}\begin{flushright}  CFTP/19-029 \end{flushright} \end{minipage}

\begin{center}

\vspace{0.50cm}

{\large \bf {Bounded masses in Two Higgs Doublets Models, Spontaneous CP Violation and $\mathbb{Z}_2$ symmetry}}\\
\vspace{0.50cm}
Miguel Nebot $^{a,b}$\footnote{\texttt{Miguel.Nebot@uv.es}}
\end{center}
%
\vspace{0.50cm}
\begin{flushleft}
\emph{$^a$ Departament de F\`\i sica Te\`orica and Instituto de F\' \i sica Corpuscular (IFIC),\\
\quad Universitat de Val\`encia -- CSIC, E-46100 Valencia, Spain.}\\
\emph{$^b$ CFTP, Instituto Superior T\' ecnico, U. de Lisboa,\\ 
\quad Av. Rovisco Pais 1, P-1049-001 Lisboa, Portugal.} 
\end{flushleft}

%
%
%
\begin{abstract}
In Two Higgs Doublet Models (2HDMs) shaped by some unbroken symmetry, imposing perturbativity requirements on the quartic couplings can imply that the allowed masses of all the fundamental scalars are bounded from above. This important property is analysed in detail for the only two realistic 2HDMs with an exact symmetry, the case with $\mathbb{Z}_2$ symmetry and the case with CP symmetry. It is also noticeable that one exception arises in each case: when the vacuum is assumed to respect the imposed symmetry, a decoupling regime can nevertheless appear without violating perturbativity requirements. In both models with an exact symmetry and no decoupling regime, soft symmetry breaking terms can however lead to a decoupling regime: the possibility that this regime might be unnatural, since it requires some fine tuning, is also analysed.
\end{abstract}

\section{Introduction\label{SEC:Introduction}}
Two Higgs-Doublets Models (2HDMs) were introduced by T.D. Lee in \cite{Lee:1973iz,Lee:1974jb}. One central and appealing motivation was the possibility that the origin of CP violation is exclusively spontaneous: with CP invariance at the Lagrangian level, CP violation can nevertheless arise from the vacuum configuration. On the other hand, a significant source of concern for 2HDMs is the presence of Scalar Flavour Changing Neutral Couplings (SFCNC): they are already present, a priori, at tree level. Safe strategies to forbid or suppress SFCNC were soon identified, like Glashow and Weinberg's Natural Flavour Conservation (NFC) \cite{Glashow:1976nt} (for recent discussions on general flavour conserving 2HDM scenarios, see \cite{Penuelas:2017ikk,Botella:2018gzy}). For 2HDMs shaped by an exact $\ZZ$ symmetry \cite{Haber:1978jt,Donoghue:1978cj,Abbott:1979dt,Hall:1981bc,Barger:1989fj,WahabElKaffas:2007xd,Aoki:2009ha} (not softly broken), this precludes a spontaneous origin of CP violation: having NFC and spontaneous CP violation (SCPV) requires more than two doublets \cite{Weinberg:1976hu,Branco:1979pv}. 2HDMs with spontaneous CP violation have been widely studied in the literature \cite{Branco:1985aq,Liu:1987ng,Asatrian:1994kk,Joshipura:2007sf,Bao:2009sa,Chen:2007nx,He:2010hz,Haber:2012np,Mao:2014oya,Mao:2016jor,Liao:2017vaf,Grzadkowski:2016szj,Ogreid:2017alh}. Recently, a 2HDM where all CP violation is originated by the vacuum, which includes SFCNC of controlled intensity, and which is viable, was presented in \cite{Nebot:2018nqn}. One important aspect of that model is the fact that the new scalars are necessarily light: their masses are all below 950 GeV. This kind of property, that the new scalars may not have arbitrarily large masses, has been noticed and explored by different authors in the context of some 2HDMs \cite{Huffel:1980sk,Casalbuoni:1987cz,Maalampi:1991fb,Kanemura:1993hm,Akeroyd:2000wc,Ginzburg:2005dt,Horejsi:2005da,Kanemura:2015ska,Flores:1982pr,Gunion:2002zf,Malinsky:2004rx,Biswas:2014uba,Bhattacharyya:2014oka,Das:2015qva,Dev:2014yca}. On that respect, it is important that the scalar potential respects boundedness from below and that the scattering of scalars at high energies is perturbatively unitary. The objective of this work is to explore the absence of such a regime with heavy new scalars (in particular the bounds on their masses) for the two viable 2HDMs with an exact symmetry, $\ZZ$ or ``standard'' CP symmetry. In the $\ZZ$ symmetric 2HDM, this property has been analyzed to some extent: it is revisited to stress some similarities with the second case, the one with ``standard'' CP symmetry, where the question is analyzed in detail here for the first time.\\
The discussion is organised as follows. Section \ref{SEC:Min:SM:2HDMSCPV} starts with the SM scalar potential and vacuum, which are briefly revisited, paying special attention to the ingredients that lead to bounds on the Higgs mass \emph{à la Lee-Quigg-Thacker} \cite{Lee:1977yc,Lee:1977eg}; decoupling and the general 2HDM are then discussed. The different symmetric 2HDMs are introduced in section \ref{SEC:2HDM:SYM}. Out of them, the only two models which are not ruled out, the one with CP symmetry and the one with $\ZZ$ symmetry, are discussed in detail. In section \ref{SEC:Num}, numerical analyses of both models are presented, showing in particular that the masses of the new scalars are constrained to be below 1 TeV. Since, as mentioned, the introduction of soft symmetry breaking terms allows the appearance of a decoupling regime, that question is addressed in section \ref{SEC:SSB}. It is stressed that, from the point of view of the symmetry, obtaining a decoupling regime is related to a rather unnatural or fine-tuned scalar potential. The appendices provide further details on different aspects of the previous sections.

\section{Minimization of the potential and absence of decoupling\label{SEC:Min:SM:2HDMSCPV}}
\subsection{Standard Model\label{sSEC:MIN:SM}}
In the Standard Model (SM), the Higgs potential is
\begin{equation}\label{eq:SM:pot:01}
\mathcal V(\SMHD)=\mu^2\SMHDd\SMHD+\lambda(\SMHDd\SMHD)^2\,,
\end{equation}
where the scalar $\SMHD$ is an $SU(2)_L$ doublet with hypercharge $Y=1/2$; boundedness from below requires $\lambda>0$.  
Electroweak symmetry is spontaneously broken (or hidden), $SU(2)_L\otimes U(1)_Y\to U(1)_Q$ with $Q=I_3+Y$, if $\mathcal V(\VEV{\SMHD})$ has a non-trivial minimum for
\begin{equation}\label{eq:SM:vev:01}
\VEV{\SMHD}=\frac{\vev{}}{\sqrt{2}}\begin{pmatrix}0\\ 1\end{pmatrix}\,.
\end{equation}
In order to have an extremum, one needs
\begin{equation}\label{eq:SM:pot:min:01}
\frac{d}{d\vev{}}\mathcal V(\VEV{\SMHD})=\vev{}(\mu^2+\lambda\vev{}^2)=0\,,
\end{equation}
that is, one needs a potential with $\mu^2=-\lambda v^2<0$. Then, the mass of the SM Higgs boson, $\nhSM$, is $\mnhSM^2=\frac{d^2}{dv^2}\mathcal V(\VEV{\SMHD})$, computed at the candidate minimum in \refEQ{eq:SM:pot:min:01}:
\begin{equation}\label{eq:SM:pot:hmass:01}
\mnhSM^2=\mu^2+3\lambda\vev{}^2=2\lambda\vev{}^2>0\,.
\end{equation}
In order to achieve the desired spontaneous symmetry breaking (that is, the correct Fermi constant $G_F$) one chooses the vacuum expectation value (vev) $\vev{}\simeq 246$ GeV. The crucial aspect is that both $\mu^2$ and, most importantly, $\mnhSM$, are fixed in terms of the vacuum expectation value $\vev{}$ and $\lambda$ (dimensionless) by means of the minimization condition.
%
Before the discovery of 2012 \cite{Aad:2012tfa,Chatrchyan:2012xdj}, one line of reasoning concerning the previous steps could be simply summarized as: \emph{any constraint on $\lambda$ translates into a constraint on $\mnhSM^2$}.\\ 
Different theoretical requirements like the stability (or metastability) of the vacuum, triviality, perturbative unitarity, were considered in order to provide, precisely, that kind of constraint  \cite{Dicus:1973vj,Weinberg:1976pe,Lee:1977yc,Lee:1977eg,Politzer:1978ic,Cabibbo:1979ay,Dashen:1983ts,Callaway:1983zd,Weldon:1984th,Casalbuoni:1986hy,Passarino:1985ax,Dawson:1988va,Durand:1989zs}. In the SM, among those constraints, the $2\to 2$ scattering of longitudinal gauge bosons and scalars at high energies depends quite straightforwardly on the coupling $\lambda$: requiring perturbative unitarity of those scattering processes gives simple bounds on $\lambda$, and, as shown by Lee, Quigg and Thacker \cite{Lee:1977yc,Lee:1977eg} (see also \cite{Dicus:1973vj}), this turns into an upper bound on $\mnhSM^2$. Of course, with the 2012 discovery, the situation is reversed for the SM Higgs: $\mnhSM$ is measured, and $\lambda$ inferred from it. The idea, however, remains an interesting possibility for extended scalar sectors, in particular 2HDMs.

\subsection{Decoupling\label{sSEC:NonDec}}
Appelquist and Carazzone presented in \cite{Appelquist:1974tg} their celebrated ``decoupling theorem'' which states that in a renormalizable theory, heavy particles with mass $M$ ``decouple'' at low energies $E\ll M$, that is, an effective theory involving only light particles is correct to order $E/M$. There are situations where the ``decoupling theorem'' does not hold (for an early discussion see \cite{Collins:1978:wz}): it is for example well known that in flavour changing neutral transitions, arising at one loop in the SM, the formal limit of large top quark mass, $m_t\to\infty$, does not suppress such processes. In this case the ``decoupling theorem'' is circumvented because in addition to power counting arguments associated to propagators, additional $m_t$ powers appear in the couplings of top quarks to longitudinal gauge bosons. This behavior is usually referred to as \emph{non-decoupling}. Interestingly, without regard to the fact that some effects of $m_t\to\infty$ are unsuppressed, if one considers perturbativity requirements on the Yukawa couplings, this limit is trivially barred.\\ 
In the following we will refer to \emph{non-decoupling} or \emph{absence of decoupling} of the new scalars ($\nH$, $\nA$ and $\cH$ in the notation of subsection \ref{sSEC:2HDM:GEN}) with respect to the SM fields, as the fact that all their masses are bounded from above once different requirements (in particular perturbativity) are considered. The absence of a high-mass regime also implies that the $E/M$ corrections of the original decoupling theorem are not arbitrarily suppressed.

\subsection{General 2HDM\label{sSEC:2HDM:GEN}}
The most general 2HDM scalar potential is
\begin{multline}\label{eq:ScalarPotential:General:01}
\mathcal V(\HD{1},\HD{2})=
\mu_{11}^2\HDd{1}\HD{1}+\mu_{22}^2\HDd{2}\HD{2}+\left(\mu_{12}^2\HDd{1}\HD{2}+\Hc\right)\\
+\lambda_1(\HDd{1}\HD{1})^2+\lambda_2(\HDd{2}\HD{2})^2+2\lambda_3(\HDd{1}\HD{1})(\HDd{2}\HD{2})+2\lambda_4(\HDd{1}\HD{2})(\HDd{2}\HD{1})\\
+\left(\lambda_5(\HDd{1}\HD{2})^2+\Hc\right)+\left(\lambda_6(\HDd{1}\HD{1})(\HDd{1}\HD{2})+\lambda_7(\HDd{2}\HD{2})(\HDd{1}\HD{2})+\Hc\right)\,.
\end{multline}
$\mu_{11}^2$, $\mu_{22}^2$ and $\lambda_i$, $i=1$ to $4$, are real, while $\mu_{12}^2$, $\lambda_j$, $j=5,6,7$, can be complex.\\
Anticipating the more detailed analysis below, the key point is that in analogy with the SM case, one would expect that in a 2HDM where dimensionful $\mu_{ij}^2$  can be traded for dimensionless $\lambda_j$'s and vacuum expectation values through the minimization conditions, an important consequence would follow: 
 if the quartic couplings $\lambda_j$ were bounded, and that would be precisely the case when one requires perturbativity or perturbatively unitary high energy scattering, then the masses of all the scalars would be necessarily bounded from above, i.e. a decoupling regime would be absent. With $\lambda_j<\mathcal O(10)$ for a very rough estimate, new scalars masses below $\sim 1$ TeV would follow. It should be noticed that these bounds on the scalar masses have a somewhat loose nature: the precise values of the largest scalar masses that are allowed directly depend on the values used in the requirements imposed on the $\lambda_j$'s. In any case large $\lambda_j$'s signal that a description in which those fundamental scalars are the relevant degrees of freedom would not be valid anymore. Of course, having a strongly interacting scalar sector is not a problem \emph{per se}, but that is not the assumption adopted here: we concentrate on the analysis of the scenarios where the fundamental scalars in 2HDMs are the relevant degrees of freedom. Let us analyze the question in more detail.
A candidate vacuum with the desired properties for electroweak symmetry breaking has
\begin{equation}\label{eq:2HDM:EWSSB:00}
\VEV{\HD{1}}=e^{i\theta_1}\begin{pmatrix} 0\\ v_1/\sqrt{2}\end{pmatrix},\quad \VEV{\HD{2}}=e^{i\theta_2}\begin{pmatrix} 0\\ v_2/\sqrt{2}\end{pmatrix},
\end{equation}
characterized by $\vev{1},\vev{2}$, real and positive, and by $\tCP=\theta_2-\theta_1$, the relative phase between $\VEV{\HD{2}}$ and $\VEV{\HD{1}}$, which is a potential source of CP violation\footnote{With no loss of generality, one can set $\theta_1=0$ in \refEQ{eq:2HDM:EWSSB:00}.}. $\{\vev{1},\vev{2}\}$ encode the same information as $\vev{}\equiv\sqrt{\vev{1}^2+\vev{2}^2}$ (which is of course chosen to be $\vev{}\simeq 246$ GeV) and $\tb\equiv\tan\beta$, $\beta\in[0;\pi/2]$, with $\cb=\cos\beta\equiv v_1/v$, $\sb=\sin\beta\equiv v_2/v$  (in the following, the compact notation $c_x\equiv\cos x$, $s_x\equiv\sin x$ is used). 
Consider now $V(\vev{1},\vev{2},\tCP)\equiv\mathcal V(\VEV{\HD{1}},\VEV{\HD{2}})$: 
\begin{multline}\label{eq:ScalarPotential:General:02}
V(\vev{1},\vev{2},\tCP)=
\mu_{11}^2\frac{\vev{1}^2}{2}+\mu_{22}^2\frac{\vev{2}^2}{2}+\re{\bar\mu_{12}^2}\vev{1}\vev{2}+\lambda_1\frac{\vev{1}^4}{4}+\lambda_2\frac{\vev{2}^4}{4}\\
+\left(\lambda_3+\lambda_4+\re{\bar\lambda_5}\right)\frac{\vev{1}^2\vev{2}^2}{2}+\re{\bar\lambda_6}\frac{\vev{1}^3\vev{2}}{2}+\re{\bar\lambda_7}\frac{\vev{1}\vev{2}^3}{2}\,,
\end{multline}
where the $\tCP$ dependence is encoded in
\begin{equation}\label{eq:ScalarPotential:General:03}
\bar\mu_{12}^2=\mu_{12}^2e^{i\tCP},\quad \bar\lambda_5=\lambda_5e^{i2\tCP},\quad\bar\lambda_6=\lambda_6e^{i\tCP},\quad \bar\lambda_7=\lambda_7e^{i\tCP}.
\end{equation}
There are \emph{three} stationarity conditions
\begin{equation}\label{eq:2HDM:MinCond:00}
\frac{\partial V}{\partial\vev{1}}=\frac{\partial V}{\partial\vev{2}}=\frac{\partial V}{\partial\tCP}=0\,,
\end{equation}
which involve, linearly, the \emph{four}\footnote{Although $\im{\bar\mu_{12}^2}$ is absent from \refEQ{eq:ScalarPotential:General:02}, $\frac{\partial}{\partial\tCP}\re{\bar\mu_{12}^2}=-\im{\bar\mu_{12}^2}$.} dimensionful quantities $\{\mu_{11}^2,\mu_{22}^2,\re{\bar\mu_{12}^2},\im{\bar\mu_{12}^2}\}$. 
It is then clear that not all of them can be traded for $\lambda_j$'s and $\{\vev{1},\vev{2},\tCP\}$ and, as a consequence, one may expect that for values of the remaining dimensionful quantity much larger than $\vev{}$, large scalar masses can be obtained (without violating bounds on the $\lambda_j$'s). Conversely, in 2HDMs where there is less parametric freedom than in the general case in \refEQ{eq:ScalarPotential:General:01}, that is in 2HDMs shaped by some symmetry\footnote{Of course, a similar situation is also to be expected in models with more than two scalar doublets, see \cite{Faro:2020qyp}.}, that possibility might be absent, and bounds on the masses might be expected. Symmetric 2HDMs are addressed in the next section: for the moment, we will just analyze in simple terms what is necessary to have large masses of the new scalars in the general 2HDM. Before proceeding with the discussion, we take a small detour (until \refEQ{eq:2HDM:Min:Immu12:00}) to fix notation and introduce the physical fields and the mass terms.\\ 
In a Higgs basis $\{H_{1},H_{2}\}$ \cite{Georgi:1978ri,Donoghue:1978cj,Botella:1994cs}, 
\begin{equation}\label{eq:HiggsBasis:01}
\begin{pmatrix}H_{1}\\ H_{2}\end{pmatrix}=\HbROT\,
\begin{pmatrix}e^{-i\theta_1}\HD{1}\\ e^{-i\theta_2}\HD{2}\end{pmatrix},\quad \text{with}\quad 
\HbROT=\begin{pmatrix}\phantom{-}\cb & \sb\\ -\sb & \cb \end{pmatrix},\ \HbROTt=\HbROTinv,
\end{equation}
only one combination of $\HD{1}$ and $\HD{2}$, $H_{1}$, has a non-vanishing vacuum expectation value:
\begin{equation}\label{eq:HiggsBasis:02}
\langle H_{1}\rangle=\frac{v}{\sqrt 2}\begin{pmatrix} 0\\ 1\end{pmatrix},\quad \langle H_{2}\rangle=\begin{pmatrix} 0\\ 0\end{pmatrix}.
\end{equation}
The usual expansion of the fields around the candidate vacuum in \refEQ{eq:2HDM:EWSSB:00} is
\begin{equation}\label{eq:FieldExp:01}
\HD{j}=e^{i\theta_j}\begin{pmatrix}\varphi^+_j\\ \frac{\vev{j}+\rho_j+i\eta_j}{\sqrt 2}\end{pmatrix},\quad 
H_{1}=\begin{pmatrix} G^+\\ \frac{v+\nHH+iG^0}{\sqrt{2}}\end{pmatrix},\quad 
H_{2}=\begin{pmatrix} \cHp\\ \frac{\nHR+i\nHI}{\sqrt{2}}\end{pmatrix}.
\end{equation}
While the would-be Goldstone bosons $G^0$, $G^\pm$ and the physical charged scalar $\cH$ are readily identified, the neutral scalars $\{\nHH,\nHR,\nHI\}$ are not mass eigenstates: their mass terms read
\begin{equation}\label{eq:NeutralMass:01}
\frac{1}{2}\begin{pmatrix}\nHH& \nHR& \nHI\end{pmatrix}\ \mNSc\ \begin{pmatrix}\nHH\\ \nHR\\ \nHI\end{pmatrix}\subset\ \mathcal V(\HD{1},\HD{2})\,,
\end{equation}
with the $3\times 3$ mass matrix $\mNSc$ real and symmetric. $\mNSc$ is diagonalised with a $3\times 3$ real orthogonal matrix $\ROTmat$,
\begin{equation}\label{eq:NeutralMass:Diag:01}
\ROTmatT\,\mNSc\,\ROTmat=\text{diag}(\mnh^2,\mnH^2,\mnA^2)\,,\quad \ROTmatinv=\ROTmatT,
\end{equation}
which defines the physical neutral scalars $\{\nh,\nH,\nA\}$:
\begin{equation}\label{eq:ScalarROT:00}
\begin{pmatrix}\nh\\ \nH\\ \nA\end{pmatrix}=\ROTmatT\begin{pmatrix}\nHH\\ \nHR\\ \nHI\end{pmatrix}\,.
\end{equation}
$\nh$ is assumed to be the SM-like Higgs with $\mnh=125$ GeV (the alignment limit in which its couplings are SM-like corresponds to $\ROT{11}\to 1$).\\ 
We can now come back to the discussion of the regime with large new masses in the general 2HDM. Through \refEQS{eq:2HDM:MinCond:00} one can express $\{\mu_{11}^2,\mu_{22}^2,\im{\bar\mu_{12}^2}\}$ in terms of $\re{\bar\mu_{12}^2}$, $\lambda_j$'s and $\{\vev{1},\vev{2},\tCP\}$:
\begin{equation}\label{eq:2HDM:Min:Immu12:00}
\sbb\im{\bar\mu_{12}^2}=-\vev{}^2\sb\cb\left\{\sbb\im{\bar\lambda_5}+\cb^2\im{\bar\lambda_6}+\sb^2\im{\bar\lambda_7}\right\},
\end{equation}
\begin{equation}\label{eq:2HDM:Min:mu11:00}
\cb\mu_{11}^2=-\sb\re{\bar\mu_{12}^2}-\frac{\vev{}^2\cb}{4}\left\{\begin{matrix}4\cb^2\lambda_1+4\sb^2\left[\lambda_3+\lambda_4+\re{\bar\lambda_5}\right]\\ +3\sbb\re{\bar\lambda_6}+2\sb^2\tb\re{\bar\lambda_7}\end{matrix}\right\},
\end{equation}
\begin{equation}\label{eq:2HDM:Min:mu22:00}
\sb\mu_{22}^2=-\cb\re{\bar\mu_{12}^2}-\frac{\vev{}^2\sb}{4}\left\{\begin{matrix}4\sb^2\lambda_2+4\cb^2\left[\lambda_3+\lambda_4+\re{\bar\lambda_5}\right]\\ +2\cb^2\tbinv\re{\bar\lambda_6}+3\sbb\re{\bar\lambda_7}\end{matrix}\right\}.
\end{equation}
Using \refEQS{eq:2HDM:Min:Immu12:00}--\eqref{eq:2HDM:Min:mu22:00}, $\mNSc$ is fully expressed in terms of $\re{\bar\mu_{12}^2}$, $\lambda_j$'s and $\{\vev{1},\vev{2},\tCP\}$. For the argument here, it is sufficient to consider $\text{Tr}[\mNSc]$ (for further details on $\mNSc$, see Appendix \ref{APP:2HDM:M2}). In the mass eigenstate basis of \refEQ{eq:ScalarROT:00}, $\text{Tr}[\mNSc]=\mnh^2+\mnH^2+\mnA^2$, while on the other hand
\begin{equation}\label{eq:Gen:trM2Neutral}
\text{Tr}[\mNSc]=-2(\tb+\tbinv)\re{\bar\mu_{12}^2}+\vev{}^2\left\{\begin{matrix}2\cb^2\lambda_1+2\sb^2\lambda_2-2\re{\bar\lambda_5}\\+(\sbb-\tbinv)\re{\bar\lambda_6}+(\sbb-\tb)\re{\bar\lambda_7}\end{matrix}\right\}.
\end{equation}
Furthermore, the mass of the charged scalar is
\begin{equation}\label{eq:Gen:M2Ch:01}
\mcH^2=-(\tb+\tbinv)\re{\bar\mu_{12}^2}-\frac{\vev{}^2}{2}\left\{2[\lambda_4+\re{\bar\lambda_5}]+\tbinv\re{\bar\lambda_6}+\tb\re{\bar\lambda_7}\right\}.
\end{equation}
In \refEQS{eq:Gen:trM2Neutral}--\eqref{eq:Gen:M2Ch:01}, one can roughly identify three scenarios where $\mnH,\mnA,\mcH\gg\vev{}$ without requiring large $\lambda_j$'s.
\begin{align}\label{eq:Gen:Decoupling:1}
&(i)\ \tbinv\gg 1\quad\text{and}\quad -\tbinv\left[\re{\bar\mu_{12}^2}+\vev{}^2\re{\bar\lambda_6}/2\right]\gg\vev{}^2,\\
\nonumber
&\qquad\qquad\text{and thus }\left\{\begin{array}{l}\mu_{11}^2\simeq -\lambda_1\vev{}^2\\ \mu_{22}^2\simeq-\tbinv\left[\re{\bar\mu_{12}^2}+\vev{}^2\re{\bar\lambda_6}/2\right]\gg\vev{}^2\end{array}\right\}
\Rightarrow\mu_{22}^2\gg\abs{\mu_{11}^2},\\
\label{eq:Gen:Decoupling:2}
&(ii)\ \tb\gg 1\quad\text{and}\quad -\tb\left[\re{\bar\mu_{12}^2}+\vev{}^2\re{\bar\lambda_7}/2\right]\gg\vev{}^2,\\
\nonumber
&\qquad\qquad\text{and thus }\left\{\begin{array}{l}\mu_{11}^2\simeq -\tb\left[\re{\bar\mu_{12}^2}+\vev{}^2\re{\bar\lambda_7}/2\right]\gg\vev{}^2\\ \mu_{22}^2\simeq-\lambda_2\vev{}^2\end{array}\right\}
\Rightarrow\mu_{11}^2\gg\abs{\mu_{22}^2},\\
\label{eq:Gen:Decoupling:3}
&(iii)\ -\re{\bar\mu_{12}^2}\gg \vev{}^2\quad\text{without regard to}\ \beta.
\end{align}
%
In the last case, for $\tb\sim\tbinv\sim\mathcal O(1)$, $-\re{\bar\mu_{12}^2}\sim \mu_{11}^2\sim \mu_{22}^2$.\\ 
This simple analysis can be rephrased in terms of the scalar potential in the Higgs basis of \refEQ{eq:HiggsBasis:01}:
\begin{multline}\label{eq:ScalarPotential:General:HiggsBasis:01}
\mathcal V(\HHD{1},\HHD{2})=
M_{11}^2 \HHDd{1}\HHD{1}+M_{22}^2 \HHDd{2}\HHD{2}+\left(M_{12}^2 \HHDd{1}\HHD{2}+\Hc\right)\\
+\Lambda_1(\HHDd{1}\HHD{1})^2+\Lambda_2(\HHD{2} \HHD{2})^2+2\Lambda_3(\HHDd{1}\HHD{1})(\HHDd{2}\HHD{2})+2\Lambda_4(\HHDd{1}\HHD{2})(\HHDd{2}\HHD{1})\\
+\left(\Lambda_5(\HHDd{1}\HHD{2})^2+\Hc\right)+\left(\Lambda_6(\HHDd{1}\HHD{1})(\HHDd{1}\HHD{2})+\Lambda_7(\HHDd{2}\HHD{2})(\HHDd{1}\HHD{2})+\Hc\right)\,.
\end{multline}
$M_{11}^2$, $M_{22}^2$ and $\Lambda_i$, $i=1$ to $4$, are real, while $M_{12}^2$, $\Lambda_j$, $j=5,6,7$, can be complex.\\ 
%
Equations \eqref{eq:Gen:trM2Neutral} and \eqref{eq:Gen:M2Ch:01}, expressed in terms of the parameters in \refEQ{eq:ScalarPotential:General:HiggsBasis:01}, give:
\begin{align}
&\text{Tr}[\mNSc]=2M_{22}^2+2\vev{}^2[\Lambda_1+\Lambda_3+\Lambda_4],\\
&\mcH^2=M_{22}^2+\vev{}^2\Lambda_3\,.
\end{align}
One can easily read that $M_{22}^2\gg\vev{}^2$ is necessary to obtain $\mnH,\mnA,\mcH\gg\vev{}$: in the Higgs basis a mass term $M_{22}^2 \HHDd{2}\HHD{2}$ with large $M_{22}^2$ is necessary, and there is no apparent obstacle for that since $M_{22}^2$ does not participate in minimization conditions that relate it to bounded quartic couplings. Of course, since\footnote{See \cite{Gunion:2005ja} for general expressions relating the parameters in the scalar potential under changes of bases $\HD{i}\mapsto U_{ij}\HD{j}$, $U\in U(2)$.}
\begin{equation}
M_{22}^2=\sb^2\mu_{11}^2+\cb^2\mu_{22}^2+\sbb\re{\bar\mu_{12}^2}\,,
\end{equation}
one can substitute the stationarity conditions in \refEQS{eq:2HDM:Min:mu11:00}--\eqref{eq:2HDM:Min:mu22:00} and obtain
\begin{equation}
M_{22}^2=-(\tti)\re{\bar\mu_{12}^2}\\
-\vev{}^2\left\{\begin{matrix}
\cb^2\sb^2(\lambda_1+\lambda_2)+(1-2\cb^2\sb^2)\left(\lambda_3+\lambda_4+\re{\bar\lambda_5}\right)\\
+\frac{1}{2}\tbinv(\cb^4+3\sb^4)\re{\bar\lambda_6}+\frac{1}{2}\tb(\sb^4+3\cb^4)\re{\bar\lambda_7}
\end{matrix}\right\}.
\end{equation}
This shows that achieving $M_{22}^2\gg \vev{}^2$ might not be trivial if there are constraints like $\bar\mu_{12}^2=0$, and in fact brings us back to \refEQS{eq:Gen:Decoupling:1}--\eqref{eq:Gen:Decoupling:3}. After this considerations on the general 2HDM, we now turn to 2HDMs with symmetry.

\section{2HDM with symmetry \label{SEC:2HDM:SYM}}
There are two classes of symmetric 2HDMs. In the first class, invariance under ``Higgs family symmetries''
\begin{equation}\label{eq:Higgs:Sym:01}
\HD{j}\mapsto \mathcal U_{jk}\HD{k},\qquad \mathcal U\in U(2),
\end{equation}
leads to three different cases:
\begin{itemize}
\item $\ZZ$ symmetry, with $\HD{1}\mapsto -\HD{1}$, $\HD{2}\mapsto \HD{2}$, and
\begin{equation}\label{eq:2HDM:Z2:01}
\mu_{12}^2=0,\quad \lambda_6=\lambda_7=0,
\end{equation}
\item $U(1)$ symmetry, with $\HD{1}\mapsto e^{i\tau}\HD{1}$, $\HD{2}\mapsto \HD{2}$ ($\tau\neq 0,\pi$) and
\begin{equation}\label{eq:2HDM:U1:01}
\mu_{12}^2=0,\quad \lambda_5=\lambda_6=\lambda_7=0,
\end{equation}
\item full $U(2)$ symmetry with
\begin{equation}\label{eq:2HDM:U2:01}
\mu_{22}^2=\mu_{11}^2,\quad\mu_{12}^2=0,\quad \lambda_2=\lambda_1,\quad \lambda_4=\lambda_1-\lambda_3,\quad \lambda_5=\lambda_6=\lambda_7=0.
\end{equation}
\end{itemize}
Following the discussion of the general 2HDM, in all three cases, with $\mu_{12}^2=0$, the dimensionful $\mu_{ii}^2$ parameters can be traded for $\lambda_j$'s and vacuum expectation values, and thus bounded masses are to be expected. In the $U(1)$ and $U(2)$ cases, having global continuous symmetries, spontaneous electroweak symmetry breaking leaves a massless scalar. Introducing soft symmetry breaking terms $\mu_{11}^2\neq\mu_{22}^2$ and $\mu_{12}^2\neq 0$ can avoid the appearance of the unwanted massless scalars, and may also open the possibility of having heavier new scalars. Since the focus in this section is on realistic 2HDMs with an exact symmetry, we do not consider these $U(1)$ and $U(2)$ invariant cases further.\\ 
From the point of view of the scalar sector alone, since there is no unwanted massless scalar in the $\ZZ$ invariant case, we can have a viable model without the need to introduce soft symmetry breaking terms: the 2HDM with $\ZZ$ symmetry is discussed in subsection \ref{sSEC:2HDMsym:Z2} below.\\
The second class of symmetric 2HDMs is given by symmetry transformations of the generalized CP type \cite{Lee:1966ik}
\begin{equation}\label{eq:CP:Sym:01}
\HD{j}\mapsto \mathcal U_{jk}\HDc{k}\,.
\end{equation}
There are, again, three possibilities.
\begin{itemize}
\item Symmetry under the usual CP (also referred to as CP1),
\begin{equation}\label{eq:CP1:01}
\HD{j}\mapsto \HDc{j}\quad \text{with all}\ \mu_{ij}^2,\lambda_j\ \text{real}.
\end{equation}
\item CP2 symmetry with 
\begin{multline}\label{eq:CP2:01}
\begin{pmatrix}\HD{1}\\ \HD{2}\end{pmatrix}\mapsto \begin{pmatrix} 0 & 1\\ -1 & 0\end{pmatrix}\begin{pmatrix}\HDc{1}\\ \HDc{2}\end{pmatrix}\quad\text{and}\\
\mu_{22}^2=\mu_{11}^2,\quad\mu_{12}^2=0,\quad \lambda_2=\lambda_1,\quad \lambda_7=-\lambda_6.
\end{multline}
\item CP3 symmetry with 
\begin{multline}\label{eq:CP3:01}
\begin{pmatrix}\HD{1}\\ \HD{2}\end{pmatrix}\mapsto \begin{pmatrix} c_\tau & s_\tau\\ -s_\tau & c_\tau\end{pmatrix}\begin{pmatrix}\HDc{1}\\ \HDc{2}\end{pmatrix},\ 0<\tau<\pi/2, \quad\text{and}\\ 
\mu_{22}^2=\mu_{11}^2,\quad\mu_{12}^2=0,\quad \lambda_2=\lambda_1,\quad \lambda_5=\lambda_1-\lambda_3-\lambda_4\in\mathbb{R},\quad \lambda_6=\lambda_7=0.
\end{multline}
\end{itemize}
While the usual CP in \refEQ{eq:CP1:01} can be extended to the fermion sector easily (by requiring the Yukawa coupling matrices to be real), extending CP2 and CP3 to the fermion sector is much more involved. As discussed in \cite{Ferreira:2010bm}, that is not achievable for the CP2 case, which forces the presence of massless fermions, while in the CP3 case, for $\tau=\pi/3$ in \refEQ{eq:CP3:01}, a viable model could, a priori, be constructed. Unfortunately, if the symmetry is exact, there is no mixing in the fermion sector, and one needs CP3 soft breaking terms, $\mu_{22}^2\neq\mu_{11}^2$ and $\mu_{12}^2\neq 0$, to overcome that difficulty. This soft breaking can still preserve the usual CP \cite{Ferreira:2010bm}, and in that scenario one is led to a particular case of the more general ``usual CP symmetry'' scenario. Consequently, we focus on the 2HDM with usual CP symmetry, which is discussed in subsection \ref{sSEC:2HDMsym:CP}.\\
Summarizing the discussion so far, two 2HDMs with an exact symmetry, $\ZZ$ or CP, are not ruled out in principle. We analyse them in more detail in the following two subsections. Although one expects that no decoupling regime is available for the new scalars, two exceptions arise, one for each symmetry, in which the new scalar masses are not bounded.

\subsection{2HDM with $\ZZ$ symmetry\label{sSEC:2HDMsym:Z2}}
Imposing symmetry under $\HD{1}\mapsto -\HD{1}$, $\HD{2}\mapsto \HD{2}$, the general 2HDM scalar potential in \refEQ{eq:ScalarPotential:General:01} is reduced to
\begin{multline}\label{eq:ScalarPotential:Z2:01}
\mathcal V(\HD{1},\HD{2})=
\mu_{11}^2\HDd{1}\HD{1}+\mu_{22}^2\HDd{2}\HD{2}
+\lambda_1(\HDd{1}\HD{1})^2+\lambda_2(\HDd{2}\HD{2})^2\\+2\lambda_3(\HDd{1}\HD{1})(\HDd{2}\HD{2})+2\lambda_4(\HDd{1}\HD{2})(\HDd{2}\HD{1})
+\lambda_5(\HDd{1}\HD{2})^2+\lambda_5^\ast(\HDd{2}\HD{1})^2\,,
\end{multline}
with $\mu_{jj}^2\in\mathbb{R}$, $\lambda_k\in\mathbb{R}$ for $k\neq 5$. The stationarity conditions in \refEQS{eq:2HDM:Min:Immu12:00}--\eqref{eq:2HDM:Min:mu22:00} become
\begin{align}
\label{eq:2HDM:Z2:Min:Immu12:00}
&0=\vev{}^2\sbb\im{\bar\lambda_5},\\
\label{eq:2HDM:Z2:Min:mu11:00}
&\cb\mu_{11}^2=-\cb\vev{}^2\left\{\cb^2\lambda_1+\sb^2\left[\lambda_3+\lambda_4+\re{\bar\lambda_5}\right]\right\},\\
\label{eq:2HDM:Z2:Min:mu22:00}
&\sb\mu_{22}^2=-\sb\vev{}^2\left\{\sb^2\lambda_2+\cb^2\left[\lambda_3+\lambda_4+\re{\bar\lambda_5}\right]\right\}.
\end{align}
One should distinguish between the two cases $\sbb=0$ and $\sbb\neq 0$. Furthermore, a rephasing of the fields only amounts to a rephasing of $\lambda_5$ and thus, without loss of generality, one can set $\im{\lambda_5}=0$ and $\re{\lambda_5}=\lambda_5$. In that case, \refEQ{eq:ScalarPotential:Z2:01} can be written in terms of real parameters: as is well known, imposing an exact $\ZZ$, there is no CP violation in the 2HDM.

\subsubsection{Inert 2HDM\label{ssSEC:2HDMsym:Z2:inert}}
For $2\vev{1}\vev{2}=\vev{}^2\sbb=0$, that is either $\sb=0$ or $\cb=0$, the basis $\{\HD{1},\HD{2}\}$ and the Higgs basis $\{\HHD{1},\HHD{2}\}$ \emph{coincide}: $\sb=0$ or $\cb=0$ correspond to the two possible identifications $\HD{1}=\HHD{1}$ or $\HD{1}=\HHD{2}$. This 2HDM, together with a fermion sector which only couples to the scalar doublet which acquires a vacuum expectation value (owing to the $\ZZ$ symmetry), is the \emph{inert} 2HDM \cite{Deshpande:1977rw}, which provides, economically, a dark matter candidate \cite{LopezHonorez:2006gr,Ilnicka:2015jba,Belyaev:2016lok}. Then, \refEQ{eq:2HDM:Z2:Min:Immu12:00} is trivially satisfied while \refEQS{eq:2HDM:Z2:Min:mu11:00}--\eqref{eq:2HDM:Z2:Min:mu22:00} give
\begin{align}
\label{eq:Z2inert2HDM:Min:00}
& \text{for }\sb=0,\qquad 
\left\{\begin{array}{l}
\text{\refEQ{eq:2HDM:Z2:Min:mu11:00}}\Rightarrow\mu_{11}^2=-\vev{}^2\lambda_1\,,\\ 
\text{\refEQ{eq:2HDM:Z2:Min:mu22:00}}\text{ trivially satisfied, arbitrary }\mu_{22}^2,
\end{array}\right.\\
\label{eq:Z2inert2HDM:Min:01}
& \text{for }\cb=0,\qquad \left\{\begin{array}{l}
\text{\refEQ{eq:2HDM:Z2:Min:mu11:00}}\text{ trivially satisfied, arbitrary }\mu_{11}^2,\\ 
\text{\refEQ{eq:2HDM:Z2:Min:mu22:00}}\Rightarrow\mu_{22}^2=-\vev{}^2\lambda_2\,.
\end{array}\right.
\end{align}
One can now obtain
\begin{equation}\label{eq:Z2inert2HDM:NeutralMasses:00}
\mNSc=
\text{diag}\left(2\lambda\vev{}^2,\ \mu^2+\vev{}^2\left[\lambda_{3}+\lambda_{4}+\re{\bar\lambda_5}\right],\ \mu^2+\vev{}^2\left[\lambda_{3}+\lambda_{4}-\re{\bar\lambda_5}\right]\right),
\end{equation}
with
\begin{equation}\label{eq:Z2inert2HDM:Pars:00}
\lambda=\lambda_1,\ \mu^2=\mu_{22}^2\ \text{ for }\sb=0,\quad \text{and}\quad 
\lambda=\lambda_2,\ \mu^2=\mu_{11}^2\ \text{ for }\cb=0.
\end{equation}
The charged scalar mass is
\begin{equation}\label{eq:Z2inert2HDM:ChargedMass:01}
\mcH^2=\mu^2+v^2\lambda_3\,.
\end{equation}
With $\mu^2\gg\vev{}^2$, $\mnH,\mnA,\mcH\gg\vev{}$ is simply achieved; the most relevant ingredient in the inert 2HDM is the requirement of $\ZZ$ invariance \emph{in the Higgs basis}.

\subsubsection{$\ZZ$-2HDM\label{ssSEC:2HDMsym:Z2}}
For $2\vev{1}\vev{2}=\vev{}^2\sbb\neq 0$ we have the ``$\ZZ$-2HDM''. \REFEQ{eq:2HDM:Z2:Min:Immu12:00} requires $\im{\bar\lambda_5}=0$, that is $2\tCP=-\text{arg}(\lambda_5)\,[\pi]$; as mentioned after \refEQ{eq:2HDM:Z2:Min:mu22:00}, one can set $\im{\lambda_5}=0$ with a simple rephasing, in which case we simply have $\stCP=0$. Then, \refEQS{eq:2HDM:Z2:Min:mu11:00}--\eqref{eq:2HDM:Z2:Min:mu22:00} impose
\begin{equation}\label{eq:Z22HDM:Min:mu11:00}
\mu_{11}^2=-\vev{}^2\left\{\cb^2\lambda_1+\sb^2\left[\lambda_3+\lambda_4+{\bar\lambda_5}\right]\right\},
\end{equation}
\begin{equation}\label{eq:Z22HDM:Min:mu22:00}
\mu_{22}^2=-\vev{}^2\left\{\sb^2\lambda_2+\cb^2\left[\lambda_3+\lambda_4+{\bar\lambda_5}\right]\right\}.
\end{equation}
From the mass matrix of the neutral scalars in Appendix \ref{APP:2HDM:M2}, one can directly read
\begin{equation}\label{eq:Z22HDM:TrMass:01}
\mnh^2+\mnH^2=2\vev{}^2\left\{\lambda_1\cb^2+\lambda_2\sb^2\right\},\quad \mnA^2=-2\vev{}^2{\bar\lambda_5}\,,
\end{equation}
while the mass of $\cH$ is
\begin{equation}\label{eq:Z22HDM:ChargedMass:01}
\mcH^2=-\vev{}^2(\lambda_4+{\bar\lambda_5})\,.
\end{equation}
Attending to \refEQS{eq:Z22HDM:TrMass:01} and \refEQ{eq:Z22HDM:ChargedMass:01} it is clear that the scalar masses in the $\ZZ$-2HDM are bounded. We now turn to the 2HDM with CP symmetry.

\subsection{2HDM with CP Symmetry\label{sSEC:2HDMsym:CP}}
Following \refEQ{eq:CP1:01}, the 2HDM scalar potential 
\begin{multline}\label{eq:ScalarPotential:CP:01}
\mathcal V(\HD{1},\HD{2})=
\mu_{11}^2\HDd{1}\HD{1}+\mu_{22}^2\HDd{2}\HD{2}+\mu_{12}^2(\HDd{1}\HD{2}+\HDd{2}\HD{1})\\
+\lambda_1(\HDd{1}\HD{1})^2+\lambda_2(\HDd{2}\HD{2})^2+2\lambda_3(\HDd{1}\HD{1})(\HDd{2}\HD{2})+2\lambda_4(\HDd{1}\HD{2})(\HDd{2}\HD{1})\\
+\lambda_5[(\HDd{1}\HD{2})^2+(\HDd{2}\HD{1})^2]+(\lambda_6\HDd{1}\HD{1}+\lambda_7\HDd{2}\HD{2})(\HDd{1}\HD{2}+\HDd{2}\HD{1})\,,
\end{multline}
with all $\mu_{ij}^2,\lambda_j$ real, respects CP invariance, \refEQ{eq:CP1:01}.\\ 
The stationarity conditions in \refEQS{eq:2HDM:Min:Immu12:00}--\eqref{eq:2HDM:Min:mu22:00} become
\begin{align}
\label{eq:2HDM:CP:Min:Immu12:00}
&\mu_{12}^2\stCP=-\frac{\vev{}^2}{2}\left\{\sbb\lambda_5\sttCP+\cb^2\lambda_6\stCP+\sb^2\lambda_7\stCP\right\},\\
\label{eq:2HDM:CP:Min:mu11:00}
&\mu_{11}^2=-\tb\mu_{12}^2\ctCP-\frac{\vev{}^2}{4}\left\{4\cb^2\lambda_1+4\sb^2\left[\lambda_3+\lambda_4+\lambda_5\cttCP\right] +3\sbb\lambda_6\ctCP+2\sb^2\tb\lambda_7\ctCP\right\},\\
\label{eq:2HDM:CP:Min:mu22:00}
&\mu_{22}^2=-\tbinv\mu_{12}^2\ctCP-\frac{\vev{}^2}{4}\left\{4\sb^2\lambda_2+4\cb^2\left[\lambda_3+\lambda_4+\lambda_5\cttCP\right] +2\cb^2\tbinv\lambda_6\ctCP+3\sbb\lambda_7\ctCP\right\}.
\end{align}
Attending to \refEQ{eq:2HDM:CP:Min:Immu12:00}, one should now distinguish between two cases, $\stCP=0$ and $\stCP\neq 0$, that we address in turn.

\subsubsection{Real 2HDM\label{ssSEC:2HDMsym:CP:Real}}
For $\stCP=0$, \refEQ{eq:2HDM:CP:Min:Immu12:00} is fulfilled without regard to $\mu_{12}^2$, $\lambda_5$, $\lambda_6$ and $\lambda_7$. Then, \refEQS{eq:2HDM:CP:Min:mu11:00}--\eqref{eq:2HDM:CP:Min:mu22:00} simply yield \refEQS{eq:2HDM:Min:mu11:00}--\eqref{eq:2HDM:Min:mu22:00} with
\begin{equation}
\bar\mu_{12}^2\mapsto \pm\mu_{12}^2,\quad \bar\lambda_5\mapsto \lambda_{5},\quad \bar\lambda_{6}\mapsto \pm\lambda_6,\quad \bar\lambda_7\mapsto\pm\lambda_7\,.
\end{equation}
where $\pm$ corresponds to $\ctCP=\pm 1$. It follows from the discussion of section \ref{sSEC:2HDM:GEN} that in this \emph{real} 2HDM (all couplings are real and there is no vacuum CP phase) one can have $\mnH,\mnA,\mcH\gg\vev{}$. For a detailed discussion of this model (and its decoupling regime), see \cite{Gunion:2002zf,Haber:2015pua}.

\subsubsection{SCPV-2HDM\label{ssSEC:2HDMsym:SCPV}}
For $\stCP\neq 0$, we have the ``SCPV-2HDM'', which incorporates a spontaneous origin for CP Violation. The stationarity conditions, as anticipated, allow us to trade \emph{all} $\mu_{ij}^2$ for $\lambda_j$'s, $\vev{}$, $\beta$ and $\tCP$:
\begin{align}
\mu_{12}^2&=-\frac{\vev{}^2}{2}\left[4\lambda_5\cb\sb\ctCP+\lambda_6\cb^2+\lambda_7\sb^2\right],\label{eq:V:SCPV:min:mu12}\\
\mu_{11}^2&=-\vev{}^2[\lambda_1\cb^2+(\lambda_3+\lambda_4-\lambda_5)\sb^2+\lambda_6\cb\sb\ctCP],\label{eq:V:SCPV:min:mu11}\\
\mu_{22}^2&=-\vev{}^2[\lambda_2\sb^2+(\lambda_3+\lambda_4-\lambda_5)\cb^2+\lambda_7\cb\sb\ctCP].\label{eq:V:SCPV:min:mu22}
\end{align}
That is, one can choose a potential in \refEQ{eq:ScalarPotential:CP:01} where $\mu_{12}^2$, $\mu_{11}^2$ and $\mu_{22}^2$ are given in \refEQS{eq:V:SCPV:min:mu12}--\eqref{eq:V:SCPV:min:mu22}, which depend on $\lambda_j$ ($j=1$ to $7$), $\vev{}$, $\beta$ and $\tCP$. The mass of the charged scalar $\cH$ is
\begin{equation}\label{eq:ChargedMass:SCPV:01}
\mcH^2=v^2(\lambda_5-\lambda_4),
\end{equation}
and, for the neutral scalars, following appendix \ref{APP:2HDM:M2}, we have
\begin{equation}\label{eq:M2:SCPV:tr}
\text{Tr}[\mNSc]=\mnh^2+\mnH^2+\mnA^2=\vev{}^2\left\{2(\lambda_1\cb^2+\lambda_2\sb^2+\lambda_5)+(\lambda_6+\lambda_7)\sbb\ctCP\right\}\,.
\end{equation}
Equations \eqref{eq:ChargedMass:SCPV:01} and \eqref{eq:M2:SCPV:tr} show that in the SCPV-2HDM, like in the $\ZZ$-2HDM, a decoupling regime is necessarily absent if perturbativity constraints are respected. Quantitatively, the most relevant consequence is that the masses of the new scalars are forced to be, roughly, below 1 TeV, and thus phenomenologically interesting.\\

As a closing remark for this section, it is also to be noticed that the two exceptional cases where bounds on the masses can be avoided, the inert 2HDM and the real 2HDM, one for each symmetry, appear when the vacuum also respects the imposed symmetry.

\section{Analysis\label{SEC:Num}}
\begin{figure}
\begin{center}
\includegraphics[width=0.6\textwidth]{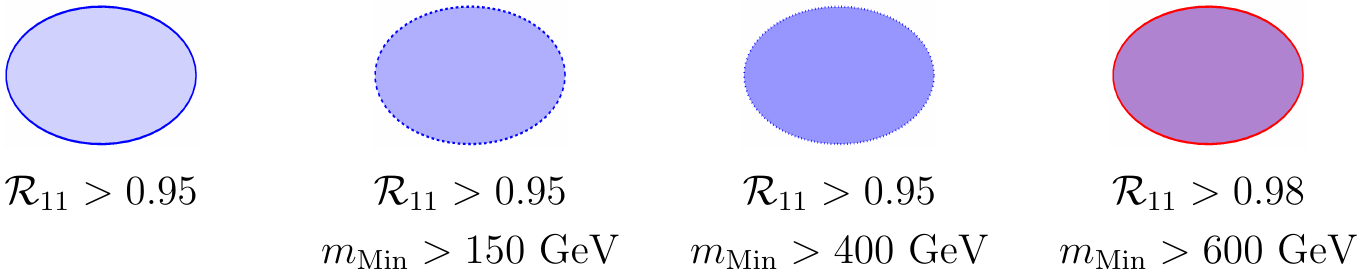}
\end{center}
\caption{Allowed regions shown in Figures \ref{FIG:SCPV:Masses:1} to \ref{FIG:mMin:ExactSyms}; they correspond to $\Delta\chi^2<3\sigma$ (for a 2D $\chi^2$ distribution) and $m_{\rm Min}\equiv\text{Min}(\mnH,\mnA,\mcH)$.\label{FIG:legend}}
\end{figure}

\begin{figure}[h!tb]
\begin{center}
\subfigure[$\mnA$ vs. $\mnH$.]{\includegraphics[width=0.3\textwidth]{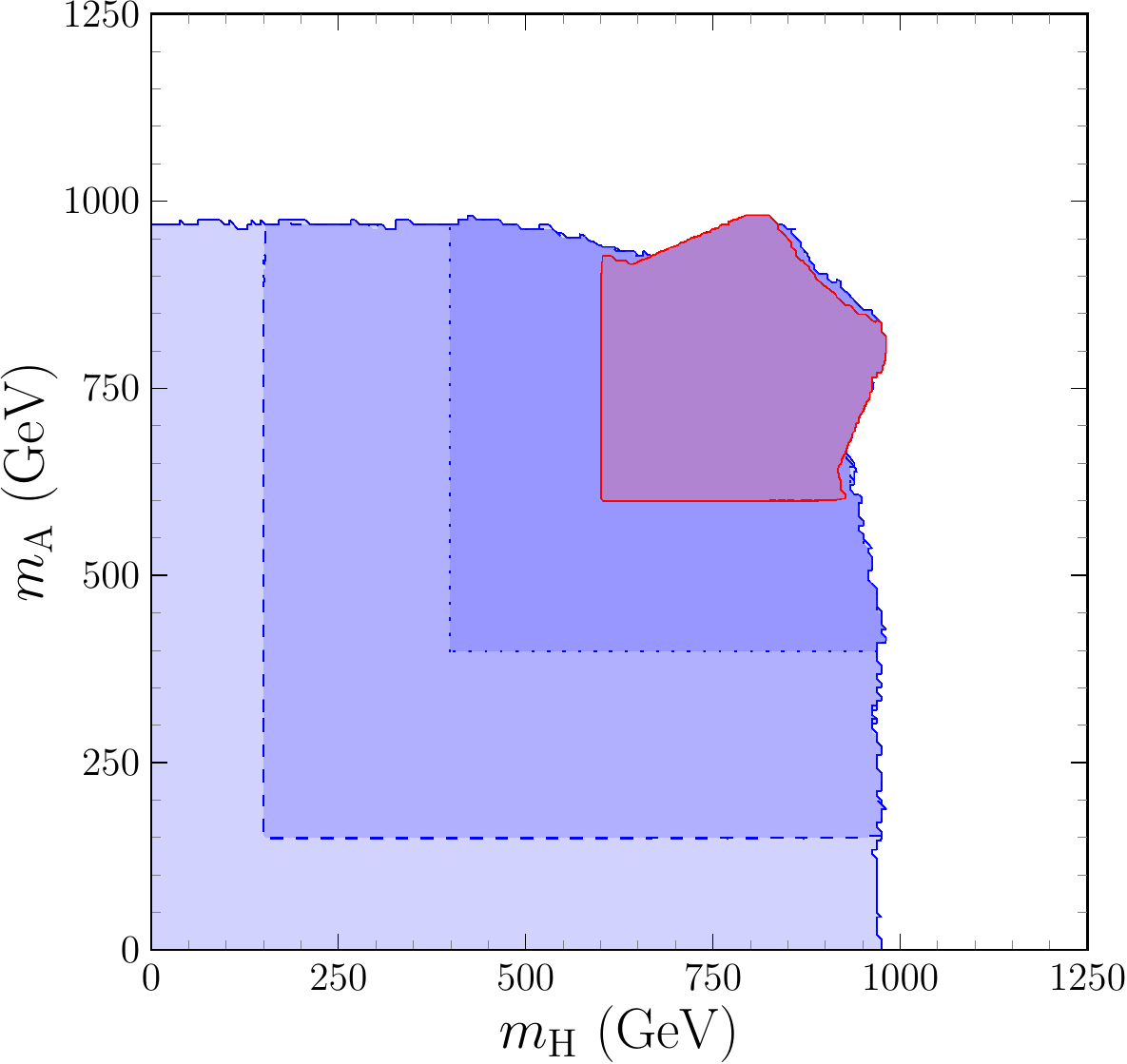}}\quad
\subfigure[$\mnA$ vs. $\mcH$.]{\includegraphics[width=0.3\textwidth]{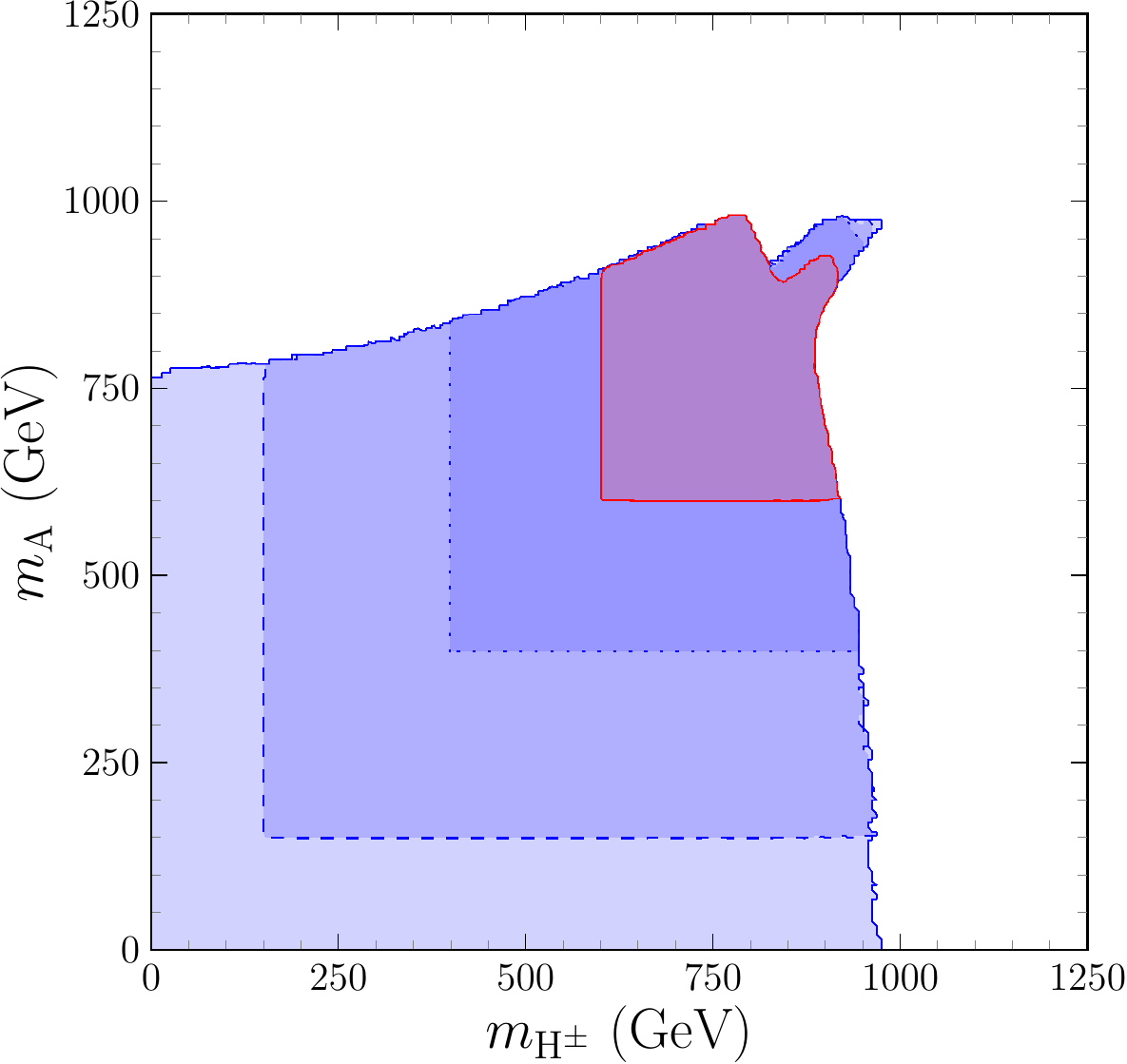}}\quad
\subfigure[$\mnH$ vs. $\mcH$.]{\includegraphics[width=0.3\textwidth]{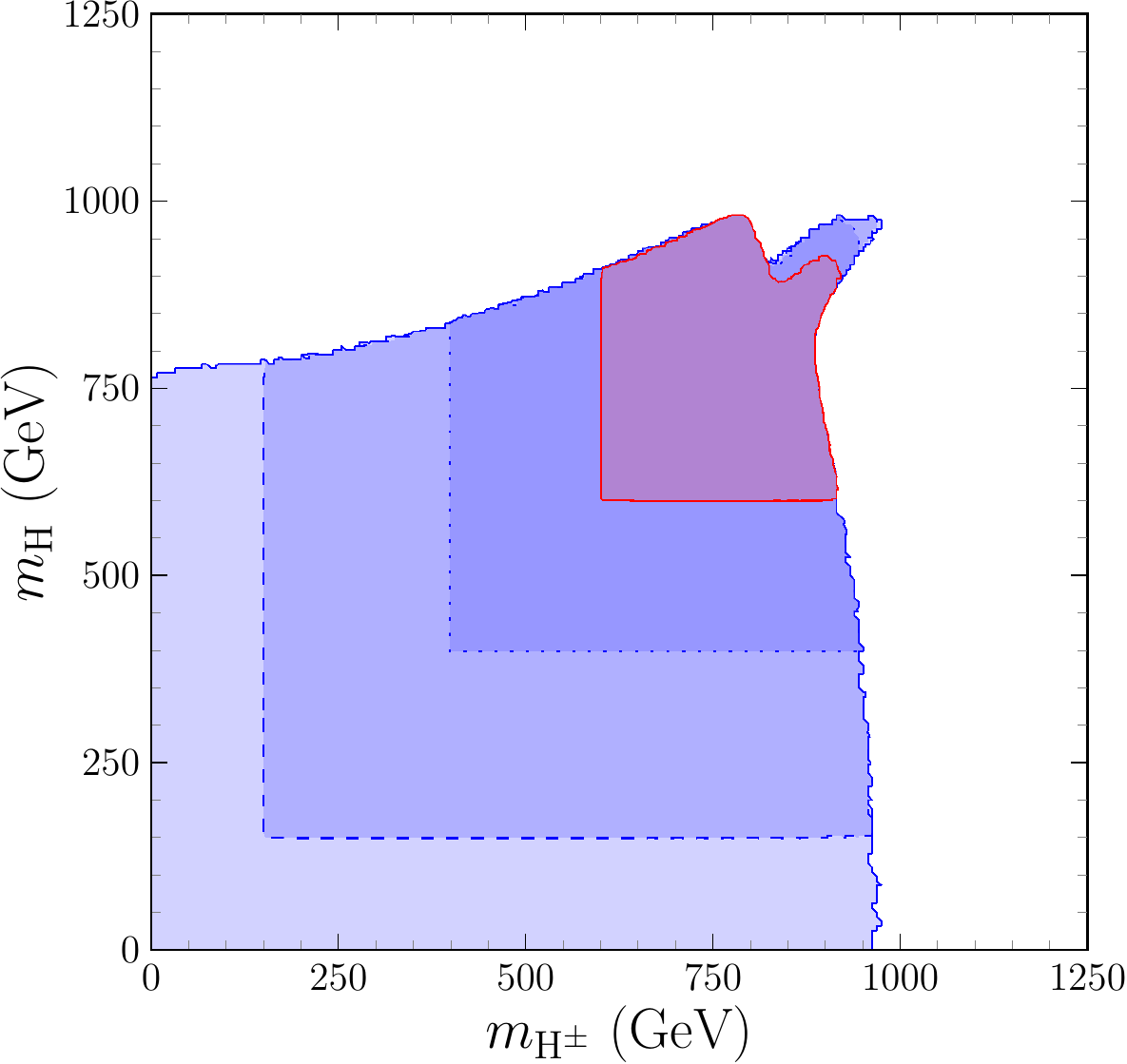}}
\end{center}
\caption{Allowed regions for the masses of the new scalars in the SCPV-2HDM (following conventions in Fig. \ref{FIG:legend}).\label{FIG:SCPV:Masses:1}}
\end{figure}

\begin{figure}[h!tb]
\begin{center}
\includegraphics[width=0.225\textwidth]{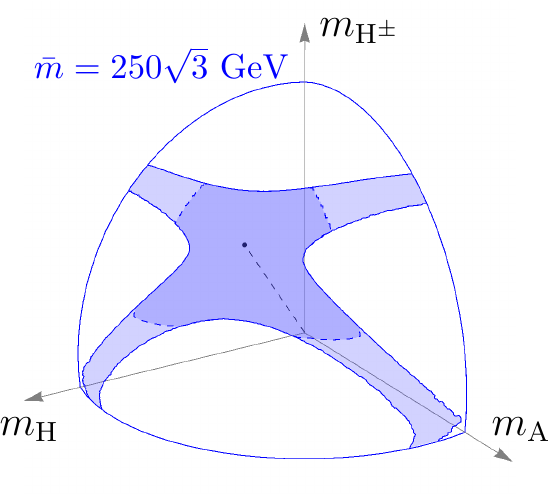}\
\includegraphics[width=0.225\textwidth]{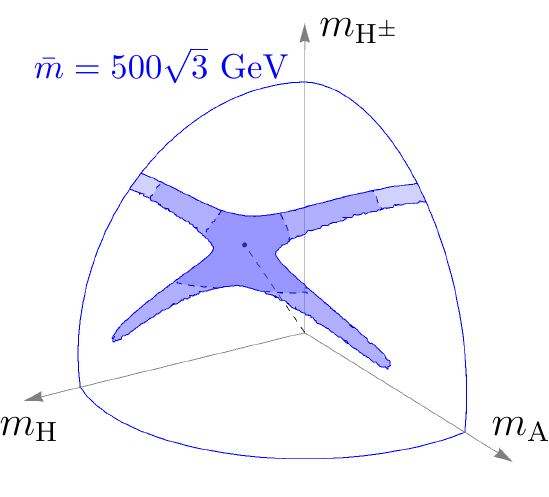}\
\includegraphics[width=0.225\textwidth]{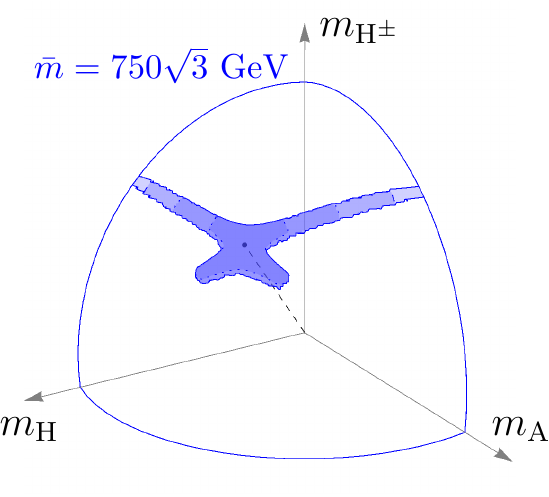}\
\includegraphics[width=0.225\textwidth]{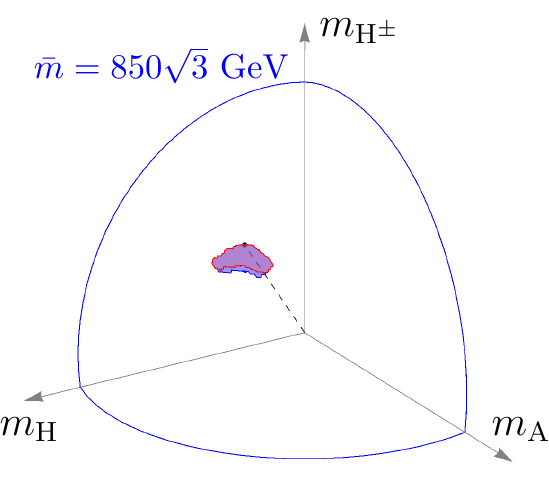}
\end{center}
\caption{Allowed regions for the masses of the new scalars in the SCPV-2HDM (following conventions in Fig. \ref{FIG:legend}), for different spherical slices in $\bar m=\sqrt{\mnH^2+\mnA^2+\mcH^2}$; the dashed straight line goes from the origin to the point $\mnH=\mnA=\mcH=\bar m/\sqrt{3}$.\label{FIG:SCPV:Masses:2}}
\end{figure}

As discussed in the previous section, there are two 2HDMs with an exact symmetry which are not ruled out by basic requirements; under simple perturbativity assumptions, they are constrained to have all new scalars relatively light. In this section we illustrate some aspects of this result through a detailed exploration of the parameter space of the models. For that exploration, the following constraints are imposed:
\begin{itemize}
\item $\mnh=125$ GeV and $\vev{}=246$ GeV (rather than a constraint, with appropriate parametrisations of the models, this simply amounts to an election of parameter values);
\item agreement with electroweak precision observables, in particular the oblique parameters $S$ and $T$ \cite{Grimus:2008nb};
\item $2\to 2$ high energy scattering is perturbatively unitary (see appendix \ref{APP:PertUnit} for details);
\item perturbativity, i.e. $\abs{\lambda_j}<4\pi$; although the high energy scattering constraint is sufficient, over most parameter space, to ensure that $\abs{\lambda_j}<4\pi$, the constraint is nevertheless imposed;
\item the scalar potential is bounded from below and the considered vacuum is the global minimum of the potential.\\
 For the $\ZZ$-2HDM, this is guaranteed by the following analytic requirements:
\begin{equation}\label{eq:2HDM:Z2:Vpos}
\lambda_1>0,\quad \lambda_2>0,\quad \sqrt{\lambda_1\lambda_2}>-\lambda_3,\quad \sqrt{\lambda_1\lambda_2}>\abs{\bar\lambda_5}-\lambda_3-\lambda_4\,,
\end{equation}
and
\begin{equation}
\left[\left(\frac{\mcH^2}{\vev{}^2}+\lambda_4\right)^2-\abs{\bar\lambda_5}^2\right]\,\left[\frac{\mcH^2}{\vev{}^2}+\sqrt{\lambda_1\lambda_2}-\lambda_3\right]>0\,.
\end{equation}
For the SCPV-2HDM, there are no simple analytic requirements as the previous, and the complete procedure described in \cite{Ivanov:2015nea} is adopted (see also \cite{Maniatis:2006fs,Kannike:2012pe}).
\end{itemize}
Since the focus is only in the scalar sector, constraints that require the specification of scalar-fermion couplings are not considered: that is the case, for example, of constraints from flavour changing transitions or from LHC production and decay processes\footnote{Although in the popular $\ZZ$ symmetric 2HDMs of types I, II (and X,Y, when the lepton sector is also considered) there is flavour conservation and all Yukawa couplings are fixed in terms of the quark masses and $\tan\beta$, that is not the case for other 2HDMs where the $\ZZ$ symmetry has a more involved realization in the fermion sector, and which have controlled SFCNC which depend on additional parameters \cite{Joshipura:1990pi,Joshipura:1990xm,Alves:2017xmk,Alves:2018kjr}.}. One cannot ignore, however, that the 125 GeV scalar is quite ``SM-like'' \cite{Khachatryan:2016vau}: in order to reflect this, a lower bound is forced on the scalar mixing element $\ROT{11}$ \cite{Bernon:2015qea,Haber:2015pua,Carena:2013ooa}. On the other hand, no direct limits are imposed on the masses of the new scalars. For better readability, the regions in Figure \ref{FIG:legend} (from left to right, each one includes the next) are shown in the plots to follow.\\ 
The allowed regions for the masses of the new scalars in the SCPV-2HDM are shown in Figure \ref{FIG:SCPV:Masses:1} (corresponding regions in the $\ZZ$-2HDM do not differ substantially). These regions trivially follow the expectations on the absence of decoupling. Figure \ref{FIG:SCPV:Masses:2} shows a different view of these allowed regions, for different ``spherical slices'' of $\bar m\equiv\sqrt{\mnH^2+\mnA^2+\mcH^2}$: notice the diminishing size of the allowed regions as $\bar m$ increases; the non-trivial shape of the regions is mainly determined by the oblique parameters $S$ and $T$; for $\bar m\sim 900\sqrt{3}$ GeV, there are no allowed regions anymore.
%

\begin{figure}[h!tb]
\begin{center}
\subfigure[$m_{\rm Min}$ vs. $\tb$ in the $\ZZ$-2HDM.\label{FIG:mMin:ExactSyms:Z2}]{\includegraphics[width=0.3377\textwidth]{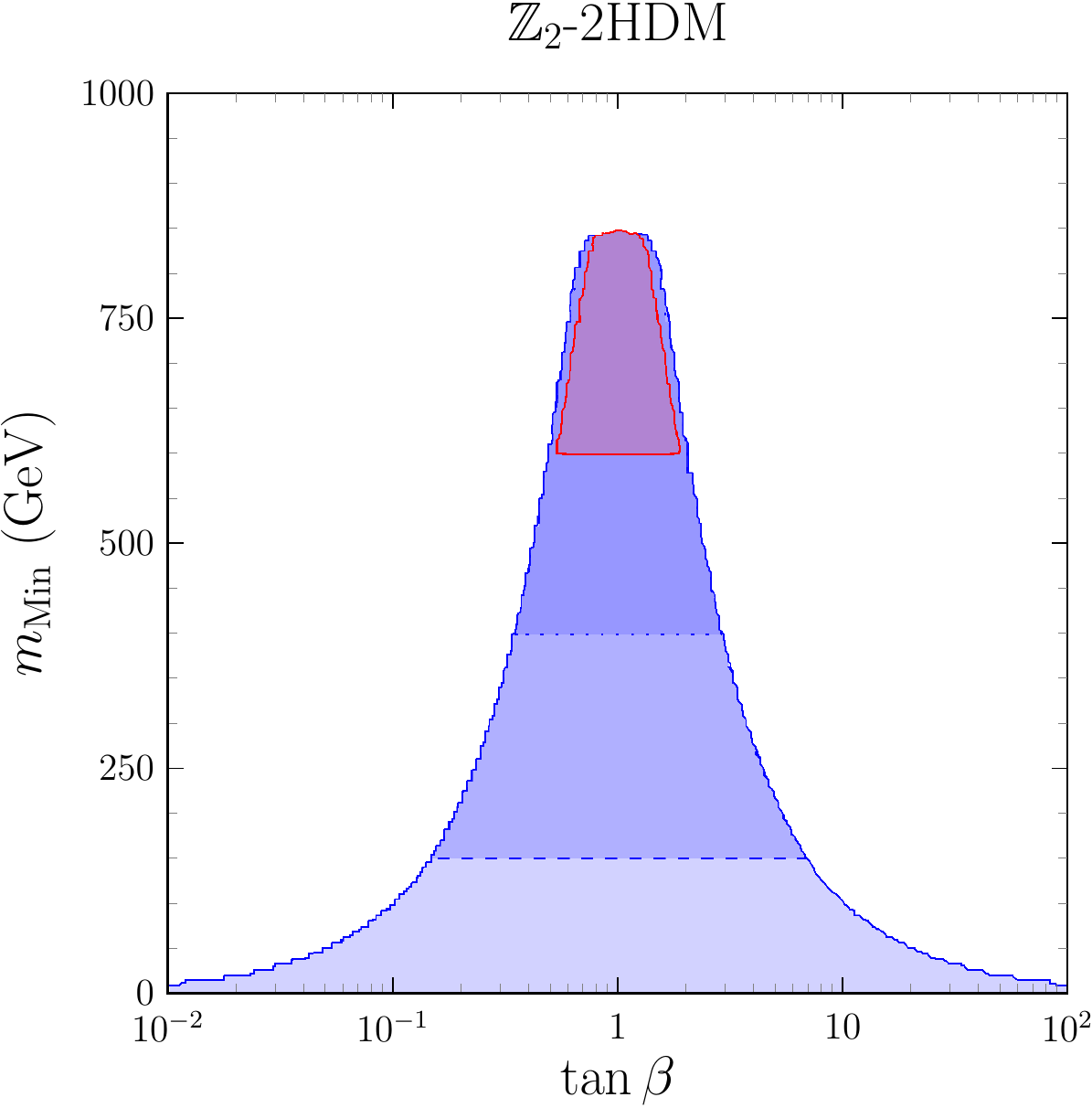}}\qquad
\subfigure[$m_{\rm Min}$ vs. $\stCP^2$ in the SCPV-2HDM.\label{FIG:mMin:ExactSyms:CP}]{\includegraphics[width=0.33\textwidth]{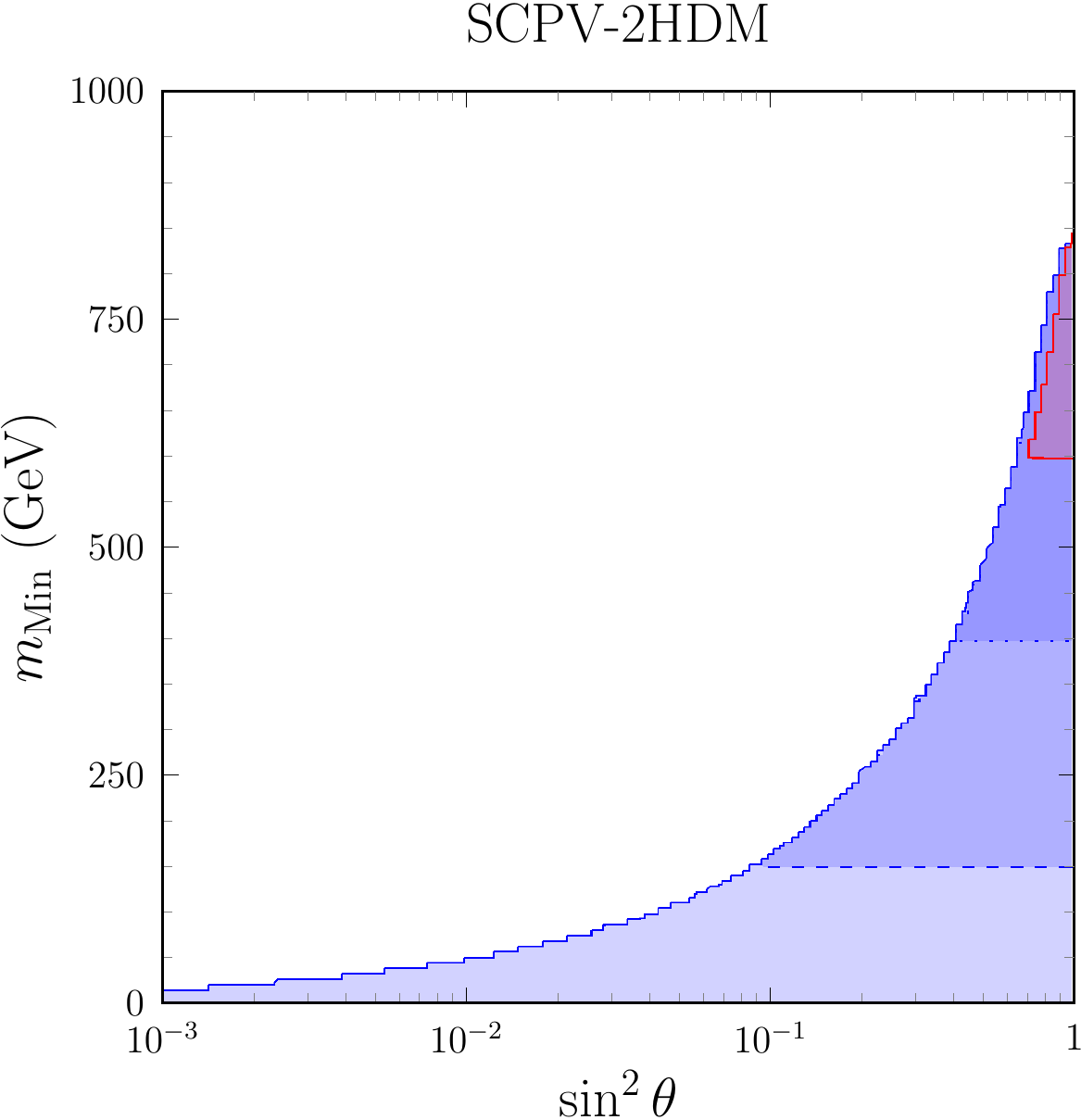}}
\end{center}
\caption{ $m_{\rm Min}$ vs. $\tb$, $\stCP^2$ (following conventions in Fig. \ref{FIG:legend}): for $\tb^{\pm 1}\to \infty$ in the $\ZZ$-2HDM, $m_{\rm Min}=\mnH\to 0$; for $\stCP\to 0$ in the SCPV-2HDM, $m_{\rm Min}=\mnA\to 0$.\label{FIG:mMin:ExactSyms}}
\end{figure}

There is also a puzzling aspect in which these numerical analyses may play a clarifying role. As discussed in section \ref{SEC:2HDM:SYM}, in the inert and the real 2HDMs, perturbativity requirements do not conflict with large scalar masses. Since one could naively expect that the real 2HDM arises in the limit $\stCP\to 0$ of the SCPV-2HDM (and similarly the inert 2HDM in the limit $\sbb\to 0$ of the $\ZZ$-2HDM), one could have accordingly expected that the allowed regions extend to $\mnH,\mnA,\mcH\gg\vev{}$ in Figures \ref{FIG:SCPV:Masses:1} and \ref{FIG:SCPV:Masses:2} in correspondence with $\stCP\to 0$. Why is that \emph{not} the case?\\
In the SCPV-2HDM, having assumed $\stCP\neq 0$, \refEQS{eq:2HDM:CP:Min:Immu12:00} and \eqref{eq:V:SCPV:min:mu12} are equivalent and give $\mu_{12}^2=-\vev{}^2[4\lambda_5\cb\sb\ctCP+\lambda_6\cb^2+\lambda_7\sb^2]/2$; on the contrary, in the real 2HDM with $\stCP=0$, \refEQ{eq:2HDM:CP:Min:Immu12:00} is trivially satisfied with free $\mu_{12}^2$. The real 2HDM \emph{is not recovered}, by construction, in the $\stCP\to 0$ limit within the SCPV-2HDM: a simple and clear manifestation of this difference appears in the mass matrices of the neutral scalars (see appendix \ref{sAPP:Masses:2HDM:SCPV}),
\begin{equation}
\text{real 2HDM: }\mNSc=\begin{pmatrix}\times & \times & 0\\ \times & \times & 0\\ 0 & 0 & \times\end{pmatrix};\qquad
\text{SCPV-2HDM with $\theta=0$: }\mathcal M_0^2=\begin{pmatrix}\times & \times & 0\\ \times & \times & 0\\ 0 & 0 & 0\end{pmatrix}.
\end{equation}
That is, the limit $\stCP\to 0$ in the SCPV-2HDM gives $\mnA\to 0$, unlike the real 2HDM.\\
For the $\ZZ$-2HDM an analogous reasoning holds. Assuming $\sbb\neq 0$, \refEQS{eq:2HDM:Z2:Min:mu11:00}--\eqref{eq:2HDM:Z2:Min:mu22:00} are equivalent to \refEQS{eq:Z22HDM:Min:mu11:00}--\eqref{eq:Z22HDM:Min:mu22:00}; however, for $\sb=0$, \refEQ{eq:2HDM:Z2:Min:mu22:00} is trivially satisfied with free $\mu_{22}^2$ while \refEQ{eq:Z22HDM:Min:mu22:00} gives $\mu_{22}^2=-\vev{}^2(\lambda_3+\lambda_4+\bar\lambda_5)$ (and similarly for $\cb\to 0$, $\mu_{11}^2$, and \refEQS{eq:Z22HDM:Min:mu11:00} and \eqref{eq:Z22HDM:Min:mu11:00}). It is then clear that the inert 2HDM \emph{is not recovered} in the limit $\sb\to 0$ ($\cb\to 0$): one cannot recover, by construction, a free $\mu_{22}^2$ ($\mu_{11}^2$) in the $\ZZ$-2HDM. A simple and clear manifestation of this difference appears again in the mass matrices of the neutral scalars (see appendix \ref{sAPP:Masses:2HDM:Z2})
\begin{equation}
\text{inert 2HDM: }\mNSc=\begin{pmatrix}\times & 0 & 0\\  0 & \times & 0\\ 0 & 0 & \times\end{pmatrix};\qquad
\text{$\mathbb{Z}_2$-2HDM with $\sbb=0$: }\mathcal M_0^2=\begin{pmatrix}\times & 0 & 0\\ 0 & 0 & 0\\ 0 & 0 & \times\end{pmatrix}.
\end{equation}
In this case, the limit $\sbb\to 0$ gives $\mnH\to 0$, unlike the inert 2HDM.\\
Figure \ref{FIG:mMin:ExactSyms} shows $m_{\rm Min}\equiv\text{Min}(\mnH,\mnA,\mcH)$ with respect to $\tb$ and $\stCP^2$ in, respectively, the $\ZZ$-2HDM and the SCPV-2HDM; not only these limits ($\sbb\to 0$ and $\stCP^2\to 0$) cannot lead to a regime with $m_{\rm Min}\gg\vev{}$, but also the largest allowed masses are only obtained in the opposite regime, $\sbb\to 1$ and $\stCP^2\to 1$.

\section{Decoupling and naturalness for softly broken $\ZZ$ or CP symmetries\label{SEC:SSB}}
As already mentioned, the introduction of soft symmetry breaking terms, that is symmetry breaking terms with mass dimension smaller than 4, possibilitates a regime where $\mnA$, $\mnH$, $\mcH\gg\vev{}$ despite perturbativity requirements. In general, an important motivation backing the introduction of soft symmetry breaking terms is the following. Since the renormalization group evolution of the soft terms (which are relevant operators), enhances them in the evolution from higher energy scales down to the electroweak scale, one can think of them, at low energies, as arising from a scenario with the symmetry almost exactly realized at high energies. Concerning decoupling of the new scalars, there is, however, a puzzling aspect: while in the model with exact symmetry the masses are bounded, when the symmetry is softly broken, a completely different qualitative regime can appear, where arbitrarily large masses are possible. In this section it is analysed that achieving a decoupling regime through soft symmetry breaking requires some tuning of the soft parameters; this tuning is not dictated by the symmetry, and can thus be interpreted as unnatural \cite{tHooft:1979rat}. 

Although fine tuning arguments have been invoked in wider contexts which include a 2HDM sector (e.g. supersymmetric extensions of the SM), there is no pretence, however, that evading the non-decoupling regime through some fine tuning constitutes a deep source of concern in the context of the 2HDMs analysed here. For a different approach to the amount of fine tuning in 2HDMs and the role of symmetries, see the detailed discussion in \cite{Draper:2016cag} (see also \cite{Haber:1989xc}).

One could object to the previous argument that, following the discussion on the Higgs basis in section \ref{sSEC:2HDM:GEN}, the regime with bounded masses is simply avoided through a large mass term for $H_2$ (the scalar doublet which does not acquire a vacuum expectation value). It would appear that there is no naturalness or fine tuning question in that case. This objection ignores, however, the presence of symmetry: in a regime with a large mass $M_{22}^2$ of $H_2$, the eventual fine tuning is already encoded in the coefficients of the scalar potential rewritten in terms of $H_1$ and $H_2$, including the mass term $M_{22}^2H_2^\dagger H_2$, as discussed at the end of section \ref{sSEC:2HDM:GEN}.

\subsection{$\ZZ$-2HDM with soft symmetry breaking\label{sSEC:SSB:Z2}}
In the $\ZZ$-2HDM, the $\ZZ$ symmetry is softly broken by adding the term $\mu_{12}^2\HDd{1}\HD{2}+\Hc$ to $\mathcal V(\HD{1},\HD{2})$ in \refEQ{eq:ScalarPotential:Z2:01}. Instead of the stationarity conditions in \refEQS{eq:2HDM:Z2:Min:Immu12:00}--\eqref{eq:2HDM:Z2:Min:mu22:00}, we now have
\begin{align}
\label{eq:2HDM:SSBZ2:Min:Immu12:00}
&\im{\bar\mu_{12}^2}=-\vev{}^2\cb\sb\im{\bar\lambda_5},\\
\label{eq:2HDM:SSBZ2:Min:mu11:00}
&\cb\mu_{11}^2=-\sb \re{\bar\mu_{12}^2}-\cb\vev{}^2\left\{\cb^2\lambda_1+\sb^2\left[\lambda_3+\lambda_4+\re{\bar\lambda_5}\right]\right\},\\
\label{eq:2HDM:SSBZ2:Min:mu22:00}
&\sb\mu_{22}^2=-\cb \re{\bar\mu_{12}^2}-\sb\vev{}^2\left\{\sb^2\lambda_2+\cb^2\left[\lambda_3+\lambda_4+\re{\bar\lambda_5}\right]\right\}.
\end{align}
Then, the mass of the charged scalar $\cH$ is
\begin{equation}\label{eq:2HDM:SSBZ2:ChargedMass:01}
\mcH^2=-(\tti)\re{\bar\mu_{12}^2}-\vev{}^2(\lambda_4+\re{\bar\lambda_5})\,,
\end{equation}
while, from the mass matrix of the neutral scalars,
\begin{equation}
\text{Tr}[\mNSc]=-2(\tti)\re{\bar\mu_{12}^2}+2\vev{}^2\left[\cb^2\lambda_1+\sb^2\lambda_2-\re{\bar\lambda_5}\right].
\end{equation}
Decoupling requires $-(\tti)\re{\bar\mu_{12}^2}\gg \vev{}^2$; then, \refEQS{eq:2HDM:SSBZ2:Min:mu11:00}--\eqref{eq:2HDM:SSBZ2:Min:mu22:00} imply
\begin{alignat}{3}\label{eq:2HDM:SSBZ2:ft}
&\text{for }\tb\sim \mathcal O(1),\quad &&\mu_{11}^2\sim\mu_{22}^2\sim -\re{\bar\mu_{12}^2}\gg\vev{}^2\,,\\
&\text{for }\tbinv\gg 1,\quad &&\mu_{11}^2\sim \vev{}^2,\ \mu_{22}^2\gg\vev{}^2\,,\\
&\text{for }\tb\gg 1,\quad &&\mu_{22}^2\sim\vev{}^2,\ \mu_{11}^2\gg\vev{}^2\,.
\end{alignat}
For $\tb\sim\mathcal O(1)$, although $\re{\bar\mu_{12}^2}$ does not respect the symmetry, it needs to be tuned to have a magnitude similar to $\mu_{11}^2$ and $\mu_{22}^2$, which do respect the symmetry.
On the other hand, for $\tb\gg 1$ or $\tbinv\gg 1$, the strong hierarchy among $\mu_{11}^2$ and $\mu_{22}^2$ (or equivalently the strong hierarchy among the vacuum expectation values) is not motivated by the symmetry and can also be interpreted as fine tuned. Furthermore, one can also notice in \refEQS{eq:2HDM:SSBZ2:Min:Immu12:00}--\eqref{eq:2HDM:SSBZ2:Min:mu22:00} that a solution with $\sb=0$ or $\cb=0$ (in correspondence with $\tbinv\gg 1$ or $\tb\gg 1$) is only strictly compatible with no symmetry breaking, $\bar\mu_{12}^2=0$, which brings us back to the inert 2HDM of section \ref{ssSEC:2HDMsym:Z2:inert}.

\subsection{SCPV-2HDM with soft symmetry breaking\label{sSEC:SSB:SCPV}}
In the SCPV-2HDM, the CP symmetry is softly broken\footnote{Through a field rephasing, one can ``move'' this symmetry violation to dimension 4 terms in $\mathcal V(\HD{1},\HD{2})$ giving also ``hard'' rather than ``soft'' breaking: since the converse, ``any hard breaking can be rephased into soft breaking'', is not true, we nevertheless maintain this abuse of language for simplicity.} for $\im{\mu_{12}^2}\neq 0$ in the scalar potential in \refEQ{eq:ScalarPotential:CP:01}. This model \cite{Ginzburg:2002wt,Ginzburg:2004vp} has been extensively explored (see for example \cite{Fontes:2017zfn}). Instead of the stationarity conditions in \refEQS{eq:2HDM:CP:Min:Immu12:00}--\eqref{eq:2HDM:CP:Min:mu22:00}, we now have
\begin{align}
\label{eq:2HDM:SSBCP:Min:Immu12:00}
&\re{\mu_{12}^2}=-\tinvtCP\im{\mu_{12}^2}-\frac{\vev{}^2}{2}\left\{2\sbb\ctCP\lambda_5+\cb^2\lambda_6+\sb^2\lambda_7\right\},\\
\label{eq:2HDM:SSBCP:Min:mu11:00}
&\cb\mu_{11}^2=\frac{\sb}{\stCP} \im{\mu_{12}^2}-\vev{}^2\cb\left\{\cb^2\lambda_1+\sb^2\left[\lambda_3+\lambda_4-\lambda_5\right]+\sb\cb\ctCP\lambda_6\right\},\\
\label{eq:2HDM:SSBCP:Min:mu22:00}
&\sb\mu_{22}^2=\frac{\cb}{\stCP} \im{\mu_{12}^2}-\vev{}^2\sb\left\{\sb^2\lambda_2+\cb^2\left[\lambda_3+\lambda_4-\lambda_5\right]+\sb\cb\ctCP\lambda_7\right\}.
\end{align}
Then, the mass of the charged scalar $\cH$ is
\begin{equation}\label{eq:2HDM:SSBCP:ChargedMass:01}
\mcH^2=\frac{\tti}{\stCP}\im{\mu_{12}^2}+\vev{}^2(\lambda_5-\lambda_4),
\end{equation}
while, from the mass matrix of the neutral scalars,
\begin{equation}\label{eq:2HDM:SSBCP:NeutralMasses:01}
\text{Tr}[\mNSc]=2\frac{\tti}{\stCP}\im{\mu_{12}^2}+2\vev{}^2\left[\cb^2\lambda_1+\sb^2\lambda_2+\lambda_5+\cb\sb\ctCP(\lambda_6+\lambda_7)\right].
\end{equation}
Decoupling requires $\frac{\tti}{\stCP}\im{\mu_{12}^2}\gg\vev{}^2$. For $\stCP\lesssim 1$, the situation is similar to the $\ZZ$-2HDM case: from \refEQS{eq:2HDM:SSBCP:Min:Immu12:00}--\eqref{eq:2HDM:SSBCP:Min:mu22:00},
\begin{alignat}{3}\label{eq:2HDM:SSBCP:ft}
\stCP\lesssim 1,\ & \text{for } \tb\sim\mathcal O(1),\quad  && \mu_{11}^2\sim\mu_{22}^2\sim\re{\mu_{12}^2}\sim\im{\mu_{12}^2}\gg\vev{}^2,\\
\stCP\lesssim 1,\ & \text{for }\tbinv\gg 1,\quad &&\mu_{11}^2\sim \vev{}^2,\ \mu_{22}^2\gg\vev{}^2,\ \re{\mu_{12}^2}\sim\im{\mu_{12}^2},\\
\stCP\lesssim 1,\ & \text{for }\tb\gg 1,\quad &&\mu_{22}^2\sim \vev{}^2,\ \mu_{11}^2\gg\vev{}^2,\ \re{\mu_{12}^2}\sim\im{\mu_{12}^2}.
\end{alignat}
The same considerations on fine tuning as in the $\ZZ$-2HDM apply.\\
For $\stCP\ll 1$, however, \refEQS{eq:2HDM:SSBCP:Min:Immu12:00}--\eqref{eq:2HDM:SSBCP:Min:mu22:00} cannot establish in general if some kind of fine tuning is necessarily present to obtain decoupling; it is to be noticed that a solution of these equations with $\stCP=0$ is only strictly compatible with no symmetry breaking, i.e. $\im{\mu_{12}^2}=0$, which brings as back to the real 2HDM of section \ref{ssSEC:2HDMsym:CP:Real}.

\subsection{Naturalness\label{sSEC:SSB:ft}}
\begin{figure}[h!tb]
\begin{center}
\includegraphics[width=0.35\textwidth]{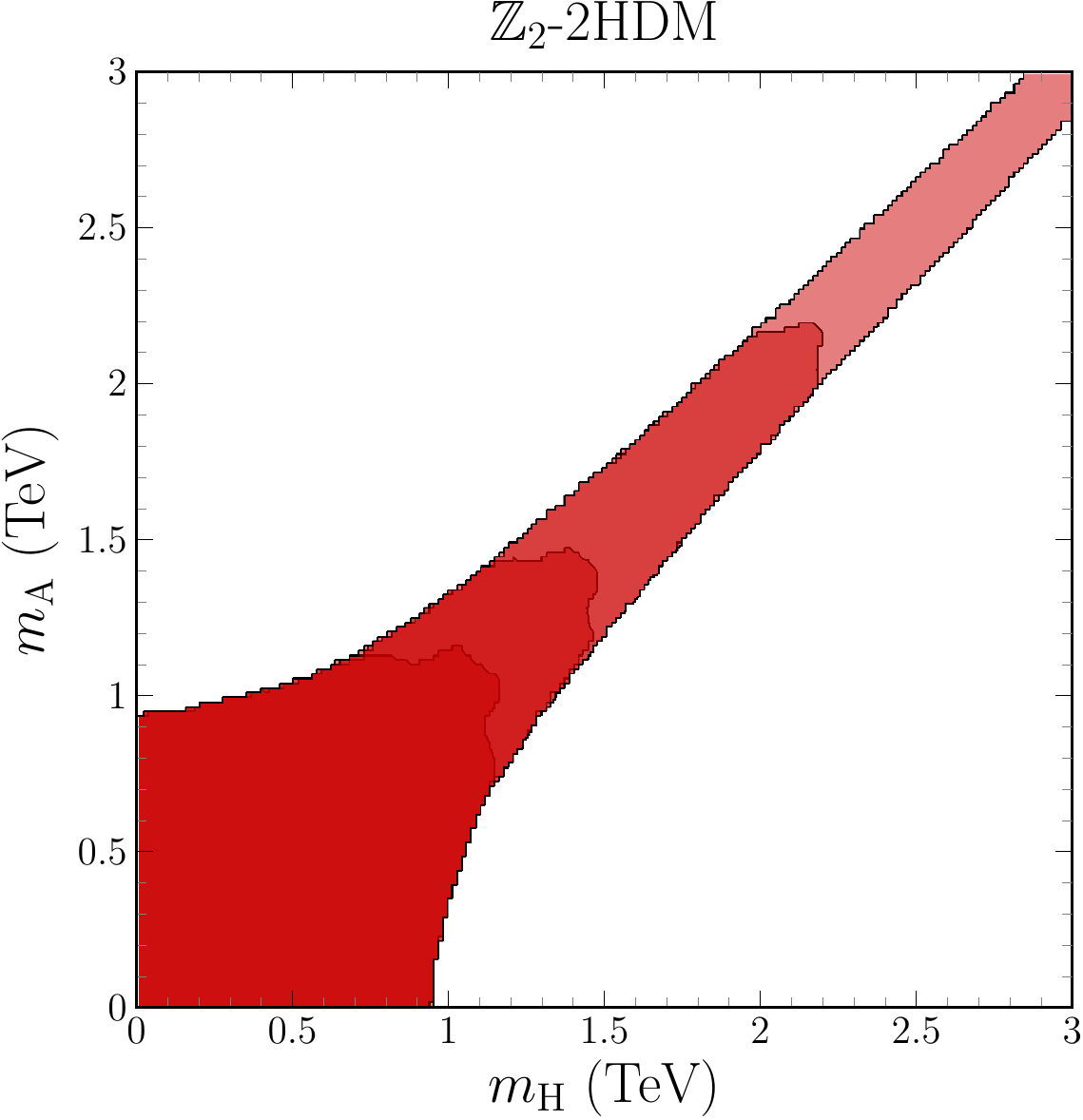}\qquad
\includegraphics[width=0.35\textwidth]{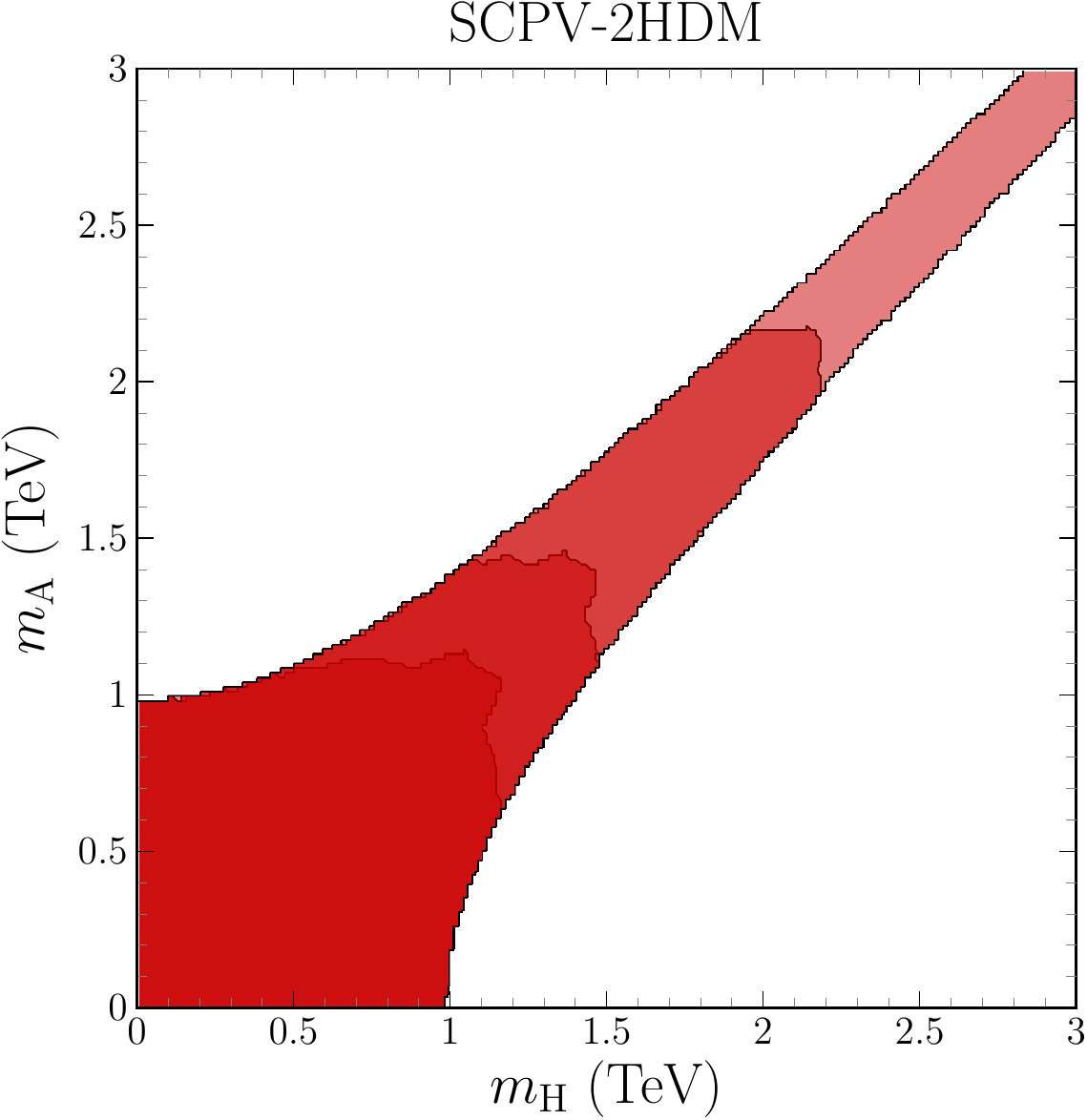}
\caption{Allowed regions for the masses of the new neutral scalars in the $\mathbb{Z}_2$-2HDM and the SCPV-2HDM with soft symmetry breaking for different fine tuning requirements: darker to lighter regions correspond to $f_{\mathbb{Z}_2},f_{\rm SCPV}>10^{-1/2},\,10^{-1},\,10^{-3/2},\,10^{-2}$.\label{FIG:ft}}
\end{center}
\end{figure}

In order to illustrate the previous discussion, one can introduce simple fine tuning measures which reflects the considerations on \refEQS{eq:2HDM:SSBZ2:ft} and on  \refEQS{eq:2HDM:SSBCP:ft}; we adopt
\begin{equation}
f_{\mathbb{Z}_2}=\frac{\vev{}^2}{{\rm Max}(|\mu_{11}^2|,|\mu_{22}^2|)},\qquad f_{\rm SCPV}=\frac{\stCP\vev{}^2}{{\rm Max}(|\mu_{11}^2|,|\mu_{22}^2|)}\,.
\end{equation}
Smaller values of $f_{\mathbb{Z}_2}$, $f_{\rm SCPV}$, correspond to larger fine tuning.
In figure \ref{FIG:ft} the allowed regions of $\mnA$ vs. $\mnH$ in the $\ZZ$-2HDM and in the SCPV-2HDM are shown for different requirements on $f_{\mathbb{Z}_2}$ and $f_{\rm SCPV}$, respectively; darker to lighter regions correspond to $f_{\mathbb{Z}_2},f_{\rm SCPV}>10^{-1/2},\,10^{-1},\,10^{-3/2},\,10^{-2}$. One can observe, for example, that masses larger than 1.5 TeV require $f_{\mathbb{Z}_2},f_{\rm SCPV}<10^{-3/2}$ while to obtain masses larger than 2 TeV $f_{\mathbb{Z}_2},f_{\rm SCPV}<10^{-2}$.

\section*{Conclusions\label{SEC:Conclusions}}
In this work, the possibility that perturbativity requirements on the quartic couplings of a 2HDM could imply that all the new scalars cannot have large masses, is analysed. We show how a decoupling regime is necessarily absent in the only two realistic models with an exact symmetry, the $\ZZ$-2HDM and the SCPV-2HDM. Although the origin of this behavior is common to both cases and rather trivial, only in the first case the question has been analyzed in detail in the literature. Numerical analyses illustrate the point and confirm quantitatively that in these models the new scalars necessarily have masses smaller than 1 TeV. Allowed ranges for these masses are fairly similar in both models. For these exact symmetries, it is also shown that (i) a decoupling regime is available for models with one specific vacuum configuration in each case -- the inert and the real 2HDM --, and (ii) these models with a decoupling regime cannot be obtained as a limiting case of the $\ZZ$-2HDM and the SCPV-2HDM. It is also possible to obtain a decoupling regime thanks to the introduction of soft symmetry breaking terms: it is finally argued that this is achieved owing to some tuning of parameters which is not justified by the symmetry, a situation that might be viewed as unnatural.

\section*{Acknowledgments}
MN thanks J.P Silva for discussions and F.J. Botella for discussions and critical reading of the manuscript. 
MN acknowledges support from Funda\c{c}\~ao para a Ci\^encia e a Tecnologia (FCT, Portugal) through projects PTDC/FIS-PAR/29436/2017 and CFTP-FCT Unit 777 (UID/FIS/00777/2013, UID/FIS/00777/2019), from Spanish grant FPA2017-85140-C3-3-P (AEI/FEDER, UE) and support from Generalitat Valenciana through CIDEGENT/2019/024.

\appendix

\section{Perturbative unitarity\label{APP:PertUnit}}
At high energies, $2\to 2$ scattering processes in the scalar sector are controlled by the quartic couplings $\lambda_j$; the corresponding $2\to 2$ tree level scattering matrix $\mathcal S$ is block diagonal, since the total hypercharge $Y$ and weak isospin $I$ are conserved in that limit  
\cite{Huffel:1980sk,Casalbuoni:1987cz,Maalampi:1991fb,Kanemura:1993hm,Akeroyd:2000wc,Ginzburg:2005dt,Horejsi:2005da,Kanemura:2015ska} (for a recent one loop analysis, see \cite{Grinstein:2015rtl}). The resulting submatrices $\mathcal S_{[Y,I]}$ are
\begin{equation}
\mathcal S_{[1,1]}=\frac{1}{8\pi}
\begin{pmatrix}
\lambda_1 & \lambda_5 & \sqrt{2}\lambda_6\\ \lambda_5^\ast & \lambda_2 & \sqrt{2}\lambda_7^\ast\\ \sqrt{2}\lambda_6^\ast & \sqrt{2}\lambda_7 & \lambda_3+\lambda_4 
\end{pmatrix}\,,
\end{equation}
\begin{equation}
\mathcal S_{[1,0]}=\frac{1}{8\pi}(\lambda_3-\lambda_4)\,,
\end{equation}
\begin{equation}
\mathcal S_{[0,1]}=\frac{1}{8\pi}
\begin{pmatrix}
\lambda_1 & \lambda_4 & \lambda_6 & \lambda_6^\ast\\ 
\lambda_4 & \lambda_2 & \lambda_7 & \lambda_7^\ast\\ 
\lambda_6^\ast & \lambda_7^\ast & \lambda_3 & \lambda_5^\ast\\
\lambda_6 & \lambda_7 & \lambda_5 & \lambda_3
\end{pmatrix}\,,
\end{equation}
\begin{equation}
\mathcal S_{[0,0]}=\frac{1}{8\pi}
\begin{pmatrix}
3\lambda_1 & 2\lambda_3+\lambda_4 & 3\lambda_6 & 3\lambda_6^\ast\\ 
2\lambda_3+\lambda_4 & 3\lambda_2 & 3\lambda_7 & 3\lambda_7^\ast\\ 
3\lambda_6^\ast & 3\lambda_7^\ast & \lambda_3+2\lambda_4 & 3\lambda_5^\ast\\
3\lambda_6 & 3\lambda_7 & 3\lambda_5 & \lambda_3+2\lambda_4
\end{pmatrix}\,.
\end{equation}
Requiring that the different $\mathcal S_{[Y,I]}$ do not yield probabilities larger than $1$ is the perturbative unitarity requirement; that is, for values of $\{\lambda_1,\lambda_2,\ldots,\lambda_7\}$ such that some eigenvalue of the above matrices is larger than $1$, that point in parameter space is not acceptable.\\
In the analyses of sections \ref{SEC:Num} and \ref{SEC:SSB}, for the $\ZZ$-2HDM one has $\lambda_6=\lambda_7=0$ and the perturbative unitarity requirement can be reformulated easily in terms of analytic conditions. For the SCPV-2HDM that is not the case, and the eigenvalues of $\mathcal S_{[Y,I]}$ are computed numerically.

\section{Mass matrices of Neutral Scalars\label{APP:2HDM:M2}}
In this appendix, the elements of the mass matrices of the neutral scalars are shown for the general 2HDM (including expressions in the Higgs basis), for the $\ZZ$-2HDM and for the SCPV-2HDM, for the $\ZZ$-2HDM with soft symmetry breaking, and for the SCPV-2HDM with soft symmetry breaking. In obtaining these mass matrices, the stationarity conditions are, of course, used; for completeness, the mass of the charged scalar $\cH$ is shown again.

\subsection{General 2HDM\label{sAPP:Masses:2HDM:gen}}
For the general 2HDM in section \ref{sSEC:2HDM:GEN}, the mass matrix of the neutral scalars is given by
\begin{align}
\nonumber
&[\mNSc]_{11}=\vev{}^2\left\{\begin{matrix}2\lambda_1\cb^4+2\lambda_2\sb^4+\left[\lambda_3+\lambda_4+\re{\bar\lambda_5}\right]\sbb^2\\ +2(\re{\bar\lambda_6}\cb^2+\re{\bar\lambda_7}\sb^2)\sbb\end{matrix}\right\},\\ 
\nonumber
& [\mNSc]_{12}=\frac{\vev{}^2}{2}\left\{\begin{matrix}2\left[-\lambda_1\cb^2+\lambda_2\sb^2\right]\sbb+\left[\lambda_3+\lambda_4+\re{\bar\lambda_5}\right]\sbN{4}\\ +\re{\bar\lambda_6}(\cbN{2}+\cbN{4})+2\re{\bar\lambda_7}\sb\sbN{3}\end{matrix}\right\},\\
\nonumber
& [\mNSc]_{22}=-(\tti)\re{\bar\mu_{12}^2}+\frac{\vev{}^2}{2}\left\{\begin{matrix}\sbb^2\left[\lambda_1+\lambda_2-2(\lambda_3+\lambda_4+\re{\bar\lambda_5})\right]\\ -(\tbinv+\sbN{4})\re{\bar\lambda_6}-(\tb-\sbN{4})\re{\bar\lambda_7}\end{matrix}\right\},\\
\nonumber
& [\mNSc]_{13}=-\vev{}^2\left\{\im{\bar\lambda_5}\sbb+\cb^2\im{\bar\lambda_6}+\sb^2\im{\bar\lambda_7}\right\},\\
\nonumber
& [\mNSc]_{23}=-\frac{\vev{}^2}{2}\left\{2\im{\bar\lambda_5}\cbb+(\im{\bar\lambda_7}-\im{\bar\lambda_6})\sbb\right\},\\
& [\mNSc]_{33}=-(\tti)\re{\bar\mu_{12}^2}-\frac{\vev{}^2}{2}\left\{4\re{\bar\lambda_5}+\tbinv\re{\bar\lambda_6}+\tb\re{\bar\lambda_7}\right\}.\label{eq:gen2HDM:M2:33}
\end{align}
The mass of the charged scalar is
\begin{equation}\label{eq:gen2HDM:MCh2:01}
\mcH^2=-(\tb+\tbinv)\re{\bar\mu_{12}^2}-\frac{\vev{}^2}{2}\left\{2[\lambda_4+\re{\bar\lambda_5}]+\tbinv\re{\bar\lambda_6}+\tb\re{\bar\lambda_7}\right\}.
\end{equation}
In terms of parameters in the Higgs basis
\begin{align}
& [\mNSc]_{11}=2\vev{}^2\Lambda_1,\nonumber\\ 
& [\mNSc]_{12}=\vev{}^2\re{\Lambda_6},\nonumber\\
& [\mNSc]_{22}=M_{22}^2+\vev{}^2\left\{\Lambda_3+\Lambda_4+\re{\Lambda_5}\right\},\nonumber\\
& [\mNSc]_{13}=-\vev{}^2\im{\Lambda_6},\nonumber\\%
& [\mNSc]_{23}=-\vev{}^2\im{\Lambda_5},\nonumber\\
& [\mNSc]_{33}=M_{22}^2+\vev{}^2\left\{\Lambda_3+\Lambda_4-\re{\Lambda_5}\right\}\,,\label{eq:gen2HDM:HiggsBasis:M2:33}
\end{align}
and
\begin{equation}\label{eq:gen2HDM:MCh2:02}
\mcH^2=M_{22}^2+\vev{}^2\Lambda_3\,.
\end{equation}
The scalar mixing matrix $\ROTmat$ in \refEQ{eq:ScalarROT:00} is a general real $3\times 3$ orthogonal matrix, which depends on three real parameters.

\subsection{$\ZZ$-2HDM\label{sAPP:Masses:2HDM:Z2}}
For the $\ZZ$-2HDM, the mass matrix of the neutral scalars is given by
\begin{align}
& [\mNSc]_{11}=2\vev{}^2\left\{\lambda_1\cb^4+\lambda_2\sb^4+2\cb^2\sb^2(\lambda_3+\lambda_4+{\bar\lambda_5})\right\},\nonumber\\ 
& [\mNSc]_{12}=\vev{}^2\sbb\left\{-\lambda_1\cb^2+\lambda_2\sb^2+\cbb(\lambda_3+\lambda_4+{\bar\lambda_5})\right\},\nonumber\\
& [\mNSc]_{22}=2\vev{}^2\cb^2\sb^2\left\{\lambda_1+\lambda_2-2(\lambda_3+\lambda_4+{\bar\lambda_5})\right\},\nonumber\\
& [\mNSc]_{13}=0,\nonumber\\%
& [\mNSc]_{23}=0,\nonumber\\
& [\mNSc]_{33}=-2\vev{}^2{\bar\lambda_5}\,.\label{eq:Z22HDM:M2:33}
\end{align}
The mass of the charged scalar is
\begin{equation}\label{eq:Z22HDM:MCh2:01}
\mcH^2=-\vev{}^2(\lambda_4+{\bar\lambda_5})\,.
\end{equation}
In this case, $\ROTmat$ is block diagonal: it is customary to introduce $\alpha$ parametrizing the transformation from $\{\rho_1,\rho_2\}$ in \refEQ{eq:FieldExp:01} to $\{\nh,\nH\}$, for example
\begin{equation}
\begin{pmatrix}\nh\\ \nH\end{pmatrix}=\begin{pmatrix}s_\alpha & c_\alpha\\ -c_\alpha& s_\alpha\end{pmatrix}\begin{pmatrix}\rho_1\\ \rho_2\end{pmatrix},
\end{equation}
and then
\begin{equation}\label{eq:Z22HDM:ROTmat}
\ROTmat=\begin{pmatrix} \sba & -\cba & 0\\ \cba & \sba & 0\\ 0 & 0 & 1\end{pmatrix}\,,
\end{equation}
which depends on a single parameter combination $\alpha+\beta$, with $\sba=\sin(\alpha+\beta)$, $\cba=\cos(\alpha+\beta)$. 

\subsection{SCPV-2HDM\label{sAPP:Masses:2HDM:SCPV}}
For the SCPV-2HDM, the mass matrix of the neutral scalars is given by
\begin{align}
\nonumber
& [\mNSc]_{11}=\vev{}^2\left\{\begin{matrix}2\lambda_1\cb^4+2\lambda_2\sb^4+\left[\lambda_3+\lambda_4+\lambda_5\cttCP\right]\sbb^2\\ +2(\lambda_6\cb^2+\lambda_7\sb^2)\sbb\ctCP\end{matrix}\right\},\\ 
\nonumber
& [\mNSc]_{12}=\vev{}^2\left\{\begin{matrix}\left[-\lambda_1\cb^2+\lambda_2\sb^2+(\lambda_3+\lambda_4+\lambda_5\cttCP)\cbb\right]\sbb\\ +\frac{1}{2}[(\lambda_6-\lambda_7)\cbN{4}+(\lambda_6+\lambda_7)\cbb]\ctCP\end{matrix}\right\},\\
\nonumber
& [\mNSc]_{22}=\vev{}^2\left\{\begin{matrix}\frac{1}{2}\sbb^2[\lambda_1+\lambda_2-2(\lambda_3+\lambda_4)]+\lambda_5(1+\cbb^2\cttCP)\\+(\lambda_7-\lambda_6)\sbb\cbb\ctCP\end{matrix}\right\},\\
\nonumber
& [\mNSc]_{13}=-\vev{}^2\stCP\left\{2\lambda_5\sbb\ctCP+\cb^2\lambda_6+\sb^2\lambda_7\right\},\\
\nonumber
& [\mNSc]_{23}=-\vev{}^2\stCP\left\{2\lambda_5\cbb\ctCP+\cb\sb(\lambda_7-\lambda_6)\right\},\\
& [\mNSc]_{33}=2\vev{}^2\lambda_5 \stCP^2\,.\label{eq:SCPV2HDM:M2:33}
\end{align}
The mass of the charged scalar is
\begin{equation}\label{eq:SCPV2HDM:MCh2:01}
\mcH^2=v^2(\lambda_5-\lambda_4)\,.
\end{equation}

\subsection{$\ZZ$-2HDM with soft symmetry breaking\label{sAPP:Masses:2HDM:Z2:SBS}}
For the $\ZZ$-2HDM with soft symmetry breaking term in section \ref{sSEC:SSB:Z2}, the mass matrix of the neutral scalars is given by
\begin{align}
& [\mNSc]_{11}=2\vev{}^2\left\{\lambda_1\cb^4+\lambda_2\sb^4+2\cb^2\sb^2(\lambda_3+\lambda_4+\re{\bar\lambda_5})\right\},\nonumber\\ 
& [\mNSc]_{12}=\vev{}^2\sbb\left\{-\lambda_1\cb^2+\lambda_2\sb^2+\cbb(\lambda_3+\lambda_4+\re{\bar\lambda_5})\right\},\nonumber\\
& [\mNSc]_{22}=-(\tti)\re{\bar\mu_{12}^2}+2\vev{}^2\cb^2\sb^2\left\{\lambda_1+\lambda_2-2(\lambda_3+\lambda_4+\re{\bar\lambda_5})\right\},\nonumber\\
& [\mNSc]_{13}=-\vev{}^2\sbb\im{\bar\lambda_5},\nonumber\\
& [\mNSc]_{23}=-\vev{}^2\cbb\im{\bar\lambda_5},\nonumber\\
& [\mNSc]_{33}=-(\tti)\re{\bar\mu_{12}^2}-2\vev{}^2\re{\bar\lambda_5}.\label{eq:2HDM:SSBZ2:M2:33}
\end{align}
The mass of the charged scalar is
\begin{equation}\label{eq:2HDM:SSBZ2:MCh2:01}
\mcH^2=-(\tti)\re{\bar\mu_{12}^2}-\vev{}^2(\lambda_4+\re{\bar\lambda_5})\,.
\end{equation}
\subsection{SCPV-2HDM with soft symmetry breaking\label{sAPP:Masses:2HDM:SCPV:SBS}}
For the SCPV-2HDM with soft symmetry breaking terms in section \ref{sSEC:SSB:SCPV}, the mass matrix of the neutral scalars is given by
\begin{align}
& [\mNSc]_{11}=\vev{}^2\left\{2\cb^4\lambda_1+2\sb^4\lambda_2+\sbb^2[\lambda_3+\lambda_4+\cttCP\lambda_5]+2\sbb\ctCP[\cb^2\lambda_6+\sb^2\lambda_7]\right\},\label{eq:SSBCP:M2:11}\\ 
& [\mNSc]_{12}=\vev{}^2\left\{\begin{array}{l}\sbb[-\cb^2\lambda_1+\sb^2\lambda_2]+\cbb\sbb[\lambda_3+\lambda_4+\cttCP\lambda_5]\\ +\ctCP[\cb\cbN{3}\lambda_6+\sb\sbN{3}\lambda_7]\end{array}\right\},\label{eq:SSBCP:M2:12}\\
& [\mNSc]_{22}=\frac{\tti}{\stCP}\im{\mu_{12}^2}+\frac{\vev{}^2}{2}\left\{\begin{array}{l}\sbb^2[\lambda_1+\lambda_2-2(\lambda_3+\lambda_4)]\\ +2(1+\cbb^2\cttCP)\lambda_5-\sbN{4}\ctCP[\lambda_6-\lambda_7]\end{array}\right\},\label{eq:SSBCP:M2:22}\\
& [\mNSc]_{13}=-\vev{}^2\stCP(2\sbb\ctCP\lambda_5+\cb^2\lambda_6+\sb^2\lambda_7),\label{eq:SSBCP:M2:13}\\%
& [\mNSc]_{23}=-\vev{}^2\stCP(2\cbb\ctCP\lambda_5-\sb\cb(\lambda_6-\lambda_7)),\label{eq:SSBCP:M2:23}\\%
& [\mNSc]_{33}=\frac{\tti}{\stCP}\im{\mu_{12}^2}+2\vev{}^2\stCP^2\lambda_5\,.\label{eq:SSBCP:M2:33}
\end{align}
The mass of the charged scalar is
\begin{equation}\label{eq:2HDM:SSBSCPV:MCh2:01}
\mcH^2=\frac{\tti}{\stCP}\im{\mu_{12}^2}+\vev{}^2(\lambda_5-\lambda_4)\,.
\end{equation}


\clearpage

\providecommand{\href}[2]{#2}\begingroup\raggedright\endgroup

\end{document}